\documentclass[11pt]{article}
\usepackage{amsmath}
\usepackage{afterpage}
\usepackage{graphicx,psfrag,epsf}
\usepackage{enumerate,color}
\usepackage{pdflscape}
\usepackage{natbib}
\usepackage{geometry}
\usepackage{rotating}
\usepackage{threeparttable}
\usepackage{url} 

\newcommand{\blind}{1}

\usepackage{fullpage}
\begin{document}

\def\spacingset#1{\renewcommand{\baselinestretch}%
{#1}\small\normalsize} \spacingset{1}


\if1\blind
{
  \title{\bf Spatial modeling of dyadic genetic relatedness data: Identifying factors associated with \textit{M. tuberculosis} transmission in Moldova}
  \date{}
  \author{Joshua L.\ Warren$^1$\thanks{This work was supported in part by the National Institute of Allergy and Infectious Diseases (R01 AI147854, R01 AI137093) and the generous support of the American people through the United States Agency for International Development (USAID) through the TREAT TB Cooperative Agreement No.\ GHN-A-00-08-00004.  The contents are the responsibility of the authors and Subgrantee and do not necessarily reflect the views of USAID or the United States Government.}, Melanie H.\ Chitwood$^2$, Benjamin Sobkowiak$^3$, \\ Valeriu Crudu$^4$, Caroline Colijn$^3$, Ted Cohen$^2$}
\date{\small $^1$Department of Biostatistics, Yale University, New Haven, CT 06510, USA\\
\small $^2$Department of Epidemiology of Microbial Diseases, Yale University, New Haven, CT 06510, USA\\
\small $^3$Department of Mathematics, Simon Fraser University, Burnaby, BC V5A 1S6, CA \\
\small $^4$Microbiology and Morphology Laboratory, Phthisiopneumology Institute, Chisinau, MLD  \\} 
  \maketitle
} \fi

\if0\blind
{
  \bigskip
  \bigskip
  \bigskip
  \begin{center}
    {\LARGE\bf Statistical methods for analyzing spatially-referenced paired genetic relatedness data}
\end{center}
  \medskip
} \fi

\bigskip
\begin{abstract}
\noindent Understanding factors that contribute to the increased likelihood of disease transmission between two individuals is important for infection control.  However, analyzing measures of genetic relatedness is complicated due to correlation arising from the presence of the same individual across multiple dyadic outcomes, potential spatial correlation caused by unmeasured transmission dynamics, and the distinctive distributional characteristics of some of the outcomes.  We develop two novel hierarchical Bayesian spatial methods for analyzing dyadic genetic relatedness data, in the form of patristic distances and transmission probabilities, that simultaneously address each of these complications.  Using individual-level spatially correlated random effect parameters, we account for multiple sources of correlation between the outcomes as well as other important features of their distribution.  Through simulation, we show the limitations of existing approaches in terms of estimating key associations of interest, and the ability of the new methodology to correct for these issues across datasets with different levels of correlation.  All methods are applied to \emph{Mycobacterium tuberculosis} data from the Republic of Moldova where we identify previously unknown factors associated with disease transmission and, through analysis of the random effect parameters, key individuals and areas with increased transmission activity.  Model comparisons show the importance of the new methodology in this setting.  The methods are implemented in the R package \texttt{GenePair}.
\end{abstract}

\noindent%
\vfill

\newpage
\spacingset{1.45} 
\section{Introduction \label{sec:intro}}
The availability and affordability of whole-genome sequencing technology has made the use of genomic data increasingly common in epidemiological modeling studies \citep{Loman, polonsky2019outbreak}.  Sequencing data from bacteriological pathogens provide insight into the spread of infectious diseases through populations and numerous methods have been developed to infer individual-to-individual transmission.  The simplest approach uses the difference of single nucleotide polymorphisms (SNPs) between two samples and assigns any pairs separated by fewer than a threshold number of SNPs to a transmission cluster \citep{RN11}.  Other approaches make use of phylogenetic trees, a reconstruction of the estimated evolutionary history of a pathogen based on genetic sequencing data \citep{Delsuc}.  For example, patristic distances, the summed length of the tree branches separating two isolates, is a similar measure of genetic distance, but this approach leverages prior biological assumptions about the substitution rate to generate a more accurate measure than SNP distance \citep{Poon}.  More complex approaches make use of transmission trees, based on phylogenetic analyses, to estimate the asymmetric probability of direct transmission between pairs of cases \citep{Didelot2017, Klinkenberg, Campbell2019}.

Understanding how potentially modifiable factors contribute to the increased likelihood of disease transmission between two individuals is important for infection control.  However, while a robust set of methods exists to estimate novel measures of genetic relatedness between case pairs, advanced statistical methods to analyze these dyadic data and their associations with other factors may require modifications/extensions due to the unique features of the outcomes (e.g., spatial correlation, distributional characteristics).  

Fixed effect regression-based approaches that do not account for correlation between the dyadic outcomes may have serious inferential limitations.  For example, we may expect correlation between dyadic outcomes $Y_{ij}$ and $Y_{kj}$, describing the genetic relatedness between individuals $i,j$ and $k,j$ respectively, because individual $j$ is present in both pairs; particularly if individual $j$ is a major driver of transmission in the population.  Ignoring this potentially positive correlation during analysis can result in overly optimistic measures of uncertainty for regression parameter estimates, leading to incorrect conclusions regarding statistical significance \citep{hoffbook}.

Several different analysis frameworks have been developed for modeling network dependence in dyadic data \citep{kenny2020dyadic}.  Random effect regression models have been widely used in this setting due to their ability to flexibly characterize the correlation, ability to handle different outcome types, and relative ease in making statistical inference \citep{warner1979new, wong1982round, gill2001statistical, hoff2002latent, hoffbook, hoff2005bilinear, hoff2021additive}.  However, current models often ignore other important sources of correlation that may be unique to the infectious disease setting (e.g., spatial correlation due to unmeasured transmission dynamics) and/or are limited in their ability to describe non-standard distributions that may be seen when analyzing novel genetic relatedness outcomes (e.g., zero-inflation), with a few exceptions.  \cite{beck2006space} and \cite{neumayer2010spatial} discuss the use of spatial lag regression models for analyzing spatially-referenced dyadic data while \cite{austin2013covariate} and \cite{ciminelli2019social} integrate spatially correlated random effects within a network analysis by extending the latent position distance model of \cite{hoff2002latent}.  With respect to non-standard distributions, \cite{simpson2015two} introduce a non-spatial (in terms of the random effect parameters) two-part mixed model for analyzing the probability of connection in whole-brain network dyadic data.

In this work, we develop hierarchical Bayesian spatial methods for analyzing dyadic genetic relatedness data in the form of patristic distances and transmission probabilities by extending the existing random effect frameworks to better reflect features of the genetic relatedness outcomes.  Specifically, instead of inducing spatial correlation in the data through the latent position distance components of the model as in \cite{austin2013covariate} and \cite{ciminelli2019social}, we incorporate spatial structure directly into the individual-level random effect parameters.  We fully investigate the implications of this choice on the induced correlation structure of the data.  Additionally, we accommodate the distinctive distributional features of the considered genetic relatedness outcomes during modeling.  Finally, we develop efficient Markov chain Monte Carlo (MCMC) sampling algorithms for several genetic relatedness outcomes, along with an R package, to facilitate posterior sampling in future work.

Using a simulation study, we show the importance of these methods for conducting accurate statistical inference for key regression associations and the limitations of existing approaches.  We also apply our methods to a unique dataset of \emph{Mycobacterium tuberculosis} isolates and associated patient data from the Republic of Moldova and show the benefits of the new methodology with respect to model fit and uncovering new insights into potentially important transmission factors.  Analyzing the posterior distributions of individual-specific random effect parameters is shown to be important for understanding how individuals in the population personally contribute to the transmission dynamics and the role of spatial versus individual variability in this process.  Given the increasing availability and interest in these types of data, and the limitations of existing approaches, we anticipate that these methods will represent important tools for researchers looking to correctly identify factors associated with genetic relatedness between individuals in future studies.  

In Section 2 we describe the data from the Republic of Moldova while the new statistical methods are presented in Section 3.  Sections 4 and 5 represent the simulation study and real data application, respectively.  We close in Section 6 with conclusions and further discussion.   

\section{Motivating data and application}
In this study, we analyze data previously described by \cite{yang2022phylogeography}.  In that work, the authors performed whole genome sequencing on \emph{Mycobacterium tuberculosis} isolates from 2,236 of the 2,770 non-incarcerated adults diagnosed with culture-positive tuberculosis (TB) in the Republic of Moldova between January 1, 2018 and December 31, 2019.  They constructed a maximum likelihood phylogenetic tree with RAxML \citep{RAxML} and identified broad, putative transmission clusters using TreeCluster \citep{TreeCluster} with a threshold of 0.001 substitutions per site. 

\begin{table}[ht!]
\centering
\caption{Summary of the dyadic genetic relatedness data in the Republic of Moldova study population by transmission probability (TP) status.  Means, with interquartile ranges given in parentheses, are shown for continuous variables and percentages are shown for categorical variables.}
\begin{tabular}{lrr}
\hline
Effect                                              & TP $= 0$ ($n=6,958)$          & TP $>0$ ($n=2,744$) \\
\hline
Distance Between Villages (km)                      &  87.49 (67.92)                & 94.76 (78.13)       \\
Same Village (\% Yes)                               &  0.56                         & 1.28                \\
Date of Diagnosis Difference (Day)                  &  225.09 (233.75)              & 202.41 (211.00)     \\
Age Difference (Year)                               &  13.57 (14.00)                & 13.33 (15.00)       \\
Sex (\%):                                           &                               &                     \\
\ \ \ Both Male                                     & 56.54                         & 58.89               \\
\ \ \ Both Female                                   &  6.11                         &  4.63               \\
\ \ \ Mixed Pair                                    & 37.35                         & 36.48               \\
Residence Location (\%):                            &                               &                     \\
\ \ \ Both Urban                                    & 14.62                         & 14.18               \\
\ \ \ Both Rural                                    & 37.55                         & 38.16               \\
\ \ \ Mixed Pair                                    & 47.83                         & 47.67               \\
Working Status (\%):                                &                               &                     \\
\ \ \ Both Employed                                 & 0.99                          & 0.77                \\
\ \ \ Both Unemployed                               & 80.81                         & 80.50               \\
\ \ \ Mixed Pair                                    & 18.19                         & 18.73               \\
Education (\%):                                     &                               &                     \\
\ \ \ Both $<$ Secondary                            & 10.30                         & 10.02               \\
\ \ \ Both $\geq$ Secondary                         & 45.33                         & 46.21               \\
\ \ \ Mixed Pair                                    & 44.37                         & 43.77               \\
\hline
\end{tabular}
\end{table}

From these analyses, \cite{yang2022phylogeography} produced estimates of genetic relatedness among sequences using two metrics.  First, they computed patristic distance between any pair of isolates within a cluster.  Patristic distance is a measure of genetic relatedness between two sequences in a phylogenetic tree expressed in substitutions per site.  Second, the authors estimated the probability that one individual infected another individual.  These probabilities are obtained using TransPhylo \citep{Didelot2017}, a Bayesian approach that augments timed phylogenetic trees with who-infected-whom information while accounting for the possibility of unsampled individuals.  TransPhylo uses an MCMC framework to obtain a posterior collection of transmission trees, accounting for the time from infection to infecting others (generation time) and the time from infection to sampling.  For most pairs $(i,j)$, there is no posterior transmission tree that includes a transmission event from $i$ to $j$ or from $j$ to $i$.  Where there are samples in the posterior in which $i$ or $j$ infected the other, the posterior probability is not necessarily symmetric.  This can occur, for example, because $i$ was sampled many months prior to $j$, making it more likely that $i$ infected $j$ than vice versa.  For all possible pairs within a transmission cluster, symmetric estimates of patristic distance (i.e., $Y_{ij} = Y_{ji}$) and asymmetric estimates of transmission probability (i.e., $Y_{ij} \neq Y_{ji}$ necessarily) are available.  To our knowledge, this is the first study to analyze transmission probabilities in the infectious disease setting.    

In addition to estimates of genetic relatedness, demographic data are available for each TB case.  These data include individual characteristics such as age (in years), sex, education status (less than secondary, secondary or higher), working status (employed, unemployed), and residence type (urban, not urban).  With these data, we calculate characteristics of the pair of individuals, such as an indicator for whether the individuals in the pair reside in the same village, the Euclidean distance between their villages of residence (in kilometers), the difference between their dates of diagnosis (in days), and the absolute difference between their ages (in years). 

\begin{figure}[ht!]
\begin{center}
\includegraphics[trim={1.95cm 0.0cm 1.00cm 0.0cm}, clip, scale = 0.38]{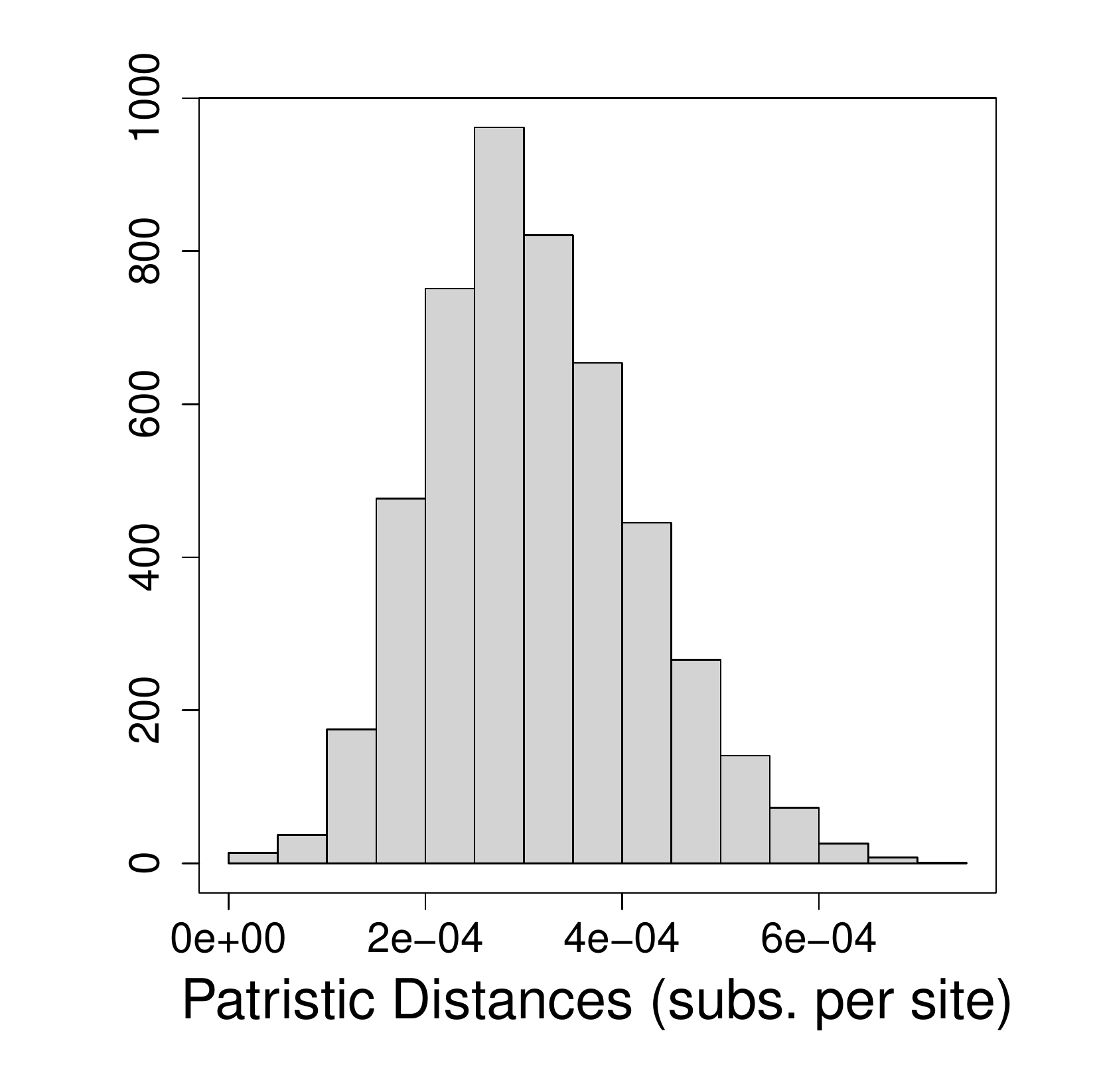}
\includegraphics[trim={1.95cm 0.0cm 1.00cm 0.0cm}, clip, scale = 0.38]{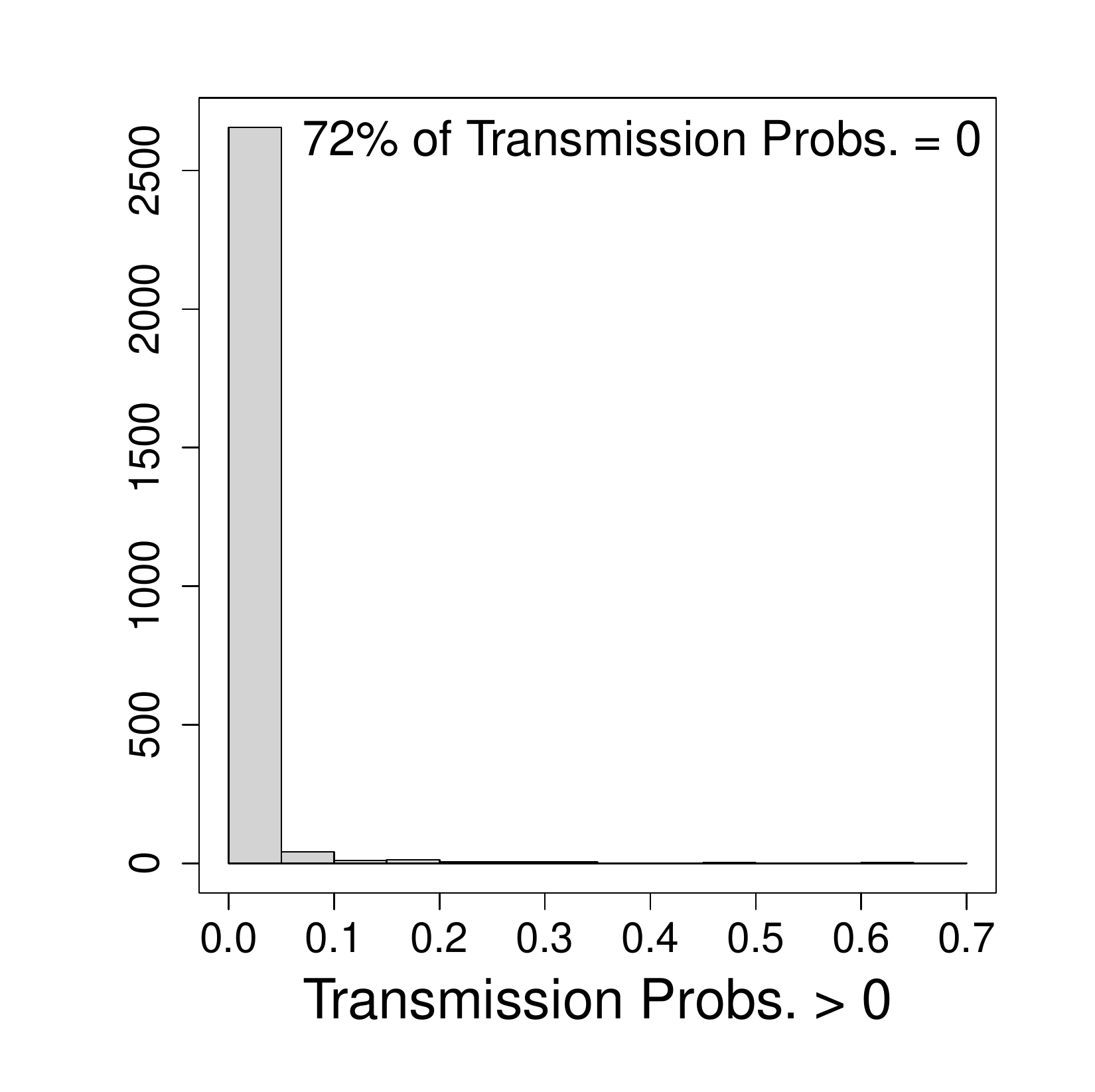}
\includegraphics[trim={1.95cm 0.0cm 0.0cm 0.0cm}, clip, scale = 0.38]{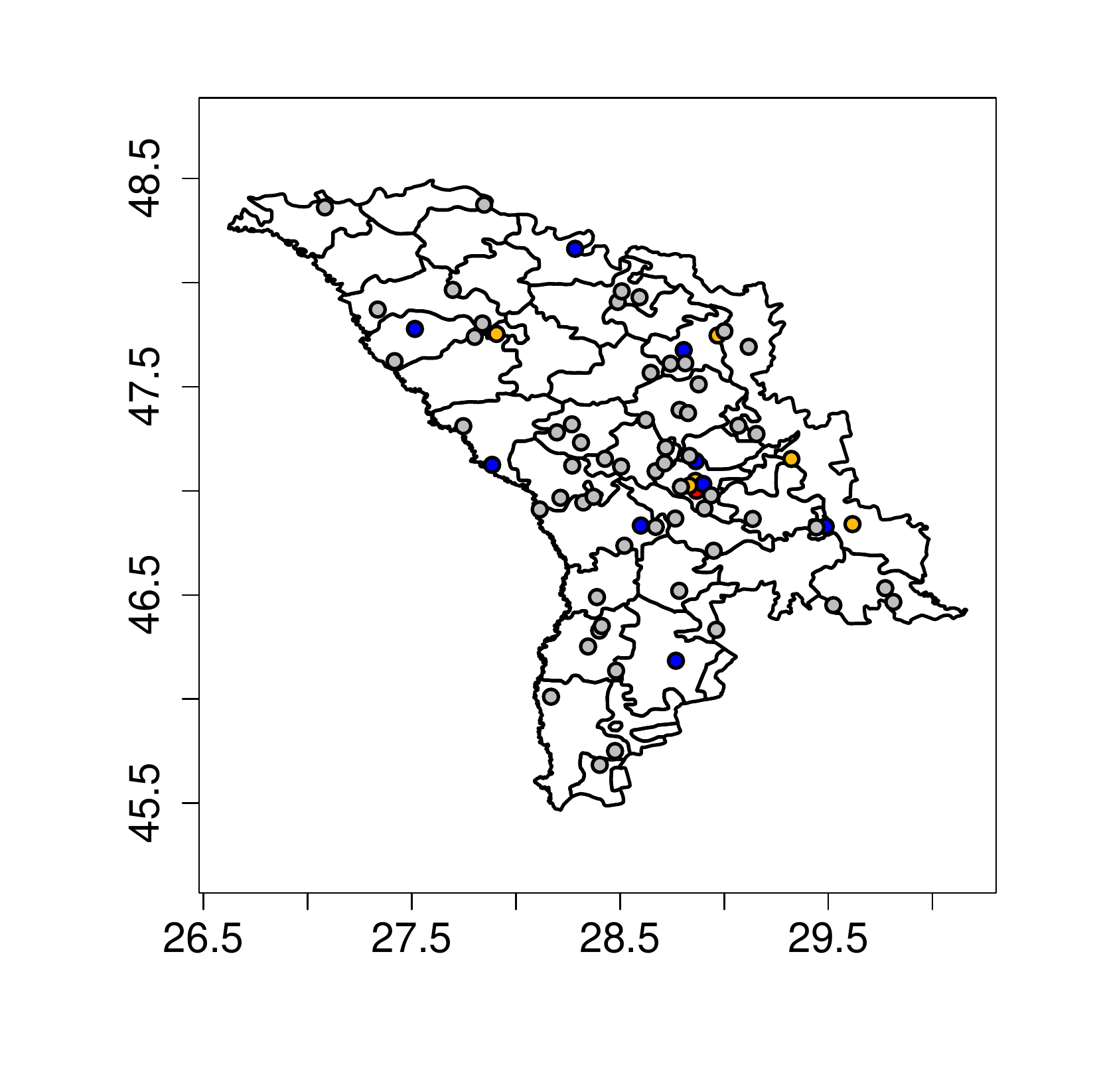}\\
\caption{Patristic distances (substitutions per site; panel 1), transmission probabilities that are $> 0$ (panel 2), and village locations (panel 3) from the largest putative cluster in the Republic of Moldova data analysis.  In panel 3, gray, blue, orange, and red points represent villages with one, two, three, and five cases, respectively.}
\end{center}
\end{figure}

In this work, we analyze data from individuals in the largest putative transmission cluster.  After removing six individuals with missing covariates, this cluster includes 99 individuals resulting in 4,851 symmetric patristic distance pairs and 9,702 asymmetric transmission probability pairs.  Characteristics of the dyadic data are shown in Table 1 while in Figure 1 we display distributions of the different genetic relatedness outcomes along with the village locations for the included individuals. 

\section{Methods}
We develop models to analyze spatially-referenced dyadic genetic relatedness outcomes (i.e., patristic distances, transmission probabilities) while accounting for multiple sources of correlation between responses and important features of the outcomes.  The methods are designed to yield accurate statistical inference for the primary regression associations of interest while also identifying individuals/locations that play a more important role in the transmission dynamics within the population.

\subsection{Patristic distances}
We model the patristic distance (log scale) between individuals $i$ and $j$ as a function of pair- and individual-level covariates as well as spatially-referenced, individual-specific random effect parameters such that \begin{equation}\ln\left(P_{ij}\right) = \textbf{x}_{ij}^{\text{T}}\boldsymbol{\beta} + \left(\textbf{d}_{i} + \textbf{d}_{j}\right)^{\text{T}}\boldsymbol{\gamma} + \theta_i + \theta_j + \epsilon_{ij},\end{equation} $i=1, \hdots, n - 1,\ j=i + 1,\hdots, n,$ where $\epsilon_{ij}|\sigma^2_{\epsilon} \stackrel{\text{iid}}{\sim} \text{N}\left(0, \sigma^2_{\epsilon}\right)$ is the error term; $n$ is the total number of individuals in the study; $P_{ij} > 0$ is the symmetric patristic distance between a unique pair of individuals $i$ and $j$ (i.e., $P_{ij} = P_{ji}$ for all $i \neq j$); $\textbf{x}_{ij}$ is a vector of covariates describing differences between individuals $i$ and $j$ (e.g., spatial distance), with $\boldsymbol{\beta}$ the corresponding vector of regression parameters; $\textbf{d}_{i}$ is a vector of covariates specific to individual $i$ where the impact of the covariates on the response, described by the $\boldsymbol{\gamma}$ vector, is assumed to be the same across all individuals; and $\theta_{i}$ is a spatially-referenced, individual-specific random effect parameter which describes individual $i$'s role in transmission events within the population (both as infector and infectee).  Small $\theta_i$ values indicate that across all outcome pairs involving individual $i$, the patristic distance is smaller on average, suggestive of an increased likelihood of being in transmission pairs.

We allow for the possibility of spatial correlation between responses by modeling the random effect parameters using a spatially-referenced Gaussian process such that \begin{align} \begin{split}
    &\theta_i = \eta\left\{h\left(\textbf{s}_i\right)\right\} + \zeta_i,\ i=1,\hdots,n, \\
    &\boldsymbol{\eta}^{\text{T}} = \left\{\eta\left(\textbf{s}_1^*\right), \hdots, \eta\left(\textbf{s}_m^*\right)\right\} | \phi, \tau^2 \sim \text{MVN}\left\{\boldsymbol{0}_m, \tau^2 \Sigma\left(\phi\right)\right\}, \text{ and}\\
    &\Sigma\left(\phi\right)_{ij} = \text{Corr}\left\{\eta\left(\textbf{s}_i^*\right), \eta\left(\textbf{s}_j^*\right)\right\} = \exp\left\{-\phi \left\|\textbf{s}_i^* - \textbf{s}_j^* \right\|\right\}
\end{split} \end{align} where $\theta_i$ is decomposed into two pieces; one that is purely spatial, $\eta\left(\textbf{s}_i^*\right)$, and one that is non-spatial, $\zeta_i$.  This allows for each parameter to be individual-specific and not driven solely by spatial location.  In the case where individuals are co-located, the function $h\left(.\right)$ maps the spatial location of an individual to an entry within a smaller set of $m < n$ unique locations such that $h\left(\textbf{s}_i\right) \in \left\{\textbf{s}_1^*, \hdots, \textbf{s}_m^*\right\}$, where it is possible that $h\left(\textbf{s}_i\right) = h\left(\textbf{s}_j\right) = \textbf{s}^*_k$ for some $i \neq j$.  When all individuals have a unique location, $h\left(\textbf{s}_i\right) = \textbf{s}^*_i$ for all $i$ and therefore, $m = n$.  

The vector of purely spatial random effect parameters, $\boldsymbol{\eta}$, is modeled using a Gaussian process centered at zero (i.e., $\textbf{0}_m$ is an $m$ length vector of zeros) with correlation structure defined by the Euclidean distances between spatial locations (i.e., $\left\|\textbf{s}^*_i - \textbf{s}^*_j\right\|$), where $\phi > 0$ controls the level of spatial correlation between the parameters and $\tau^2$ the total variability of the spatial process.  Small values of $\phi$ indicate strong spatial correlation even at larger distances.  Finally, $\zeta_i|\sigma^2_{\zeta} \stackrel{\text{iid}}{\sim}\text{N}\left(0, \sigma^2_{\zeta}\right)$ represent the individual-specific parameters that account for the possibility that two people could be in very similar spatial locations but have vastly different patterns of behavior that impact their likelihood of being transmitted to and/or transmitting to others.  
  
\subsubsection{Prior distributions}
We assign weakly informative prior distributions to the remaining model parameters when appropriate.  The regression parameters are specified as $\beta_{j}, \gamma_{k} \stackrel{\text{iid}}{\sim}\text{N}\left(0,100^2\right)$ for $j=1,\hdots,p_x$ and $k=1,\hdots,p_d$, where $p_x$ and $p_d$ are the lengths of the $\textbf{x}_{ij}$ and $\textbf{d}_i$ vectors, respectively; the variance parameters as $\sigma^2_{\epsilon}, \tau^2, \sigma^2_{\zeta} \stackrel{\text{iid}}{\sim} \text{Inverse Gamma}\left(0.01, 0.01\right)$; and the spatial correlation parameter as $\phi \sim \text{Gamma}\left(1.00, 1.00\right)$.  We scale the spatial distances used in the analysis to allow the prior distribution for $\phi$ to be minimally informative (i.e., ranging from relatively weak to strong spatial correlation at both short and long spatial distances).

\subsection{Transmission probabilities}
Next, we introduce a method for analyzing transmission probabilities which, unlike patristic distances, contain potentially rich information regarding the direction of transmission.  As a result, the outcomes are not necessarily symmetric as they were in Section 3.1 (i.e., $T_{ij} \neq T_{ji}$ necessarily).  Here, we introduce a framework for analyzing transmission probabilities that includes similar inferential goals as the model in (1, 2) while also accounting for the asymmetry of the outcome as well as other important data features (e.g., zero-inflation).

Specifically, we model the probability that individual $j$ infected individual $i$ (i.e., $T_{ij}$) as a function of individual- and pair-specific covariates and individual/location-specific random effect parameters using a mixed-type distribution.  This specification includes a binary component to account for the large proportion of exact zero transmission probabilities (see Figure 1), and a continuous piece to model the non-zero probabilities.  We define the probability density function (pdf) for a transmission probability, $f_{t_{ij}}\left(t\right)$, as \begin{equation}T_{ij} \stackrel{\text{ind}}{\sim} f_{t_{ij}}\left(t\right) = \left(1 - \pi_{ij}\right)^{1\left(t = 0\right)} \left[\frac{\pi_{ij}}{t\left(1-t\right)}f_{w_{ij}}\left\{\ln\left(\frac{t}{1-t}\right)\right\}\right]^{1\left(t > 0\right)},\ t \in \left[0,1\right)\end{equation} where all pairs are now included in the analysis (i.e., $i=1,\hdots,n$; $j=1,\hdots,n$; $i \neq j$); $1\left(.\right)$ is an indicator function taking the value of one when the input statement is true and the value of zero otherwise; the transmission probabilities can be exactly equal to zero; and $\left(1 - \pi_{ij}\right)$ describes the probability that this event occurs.  

We use a logistic regression framework to connect these underlying probabilities with covariates and random effect parameters such that \begin{equation}\ln\left(\frac{\pi_{ij}}{1 - \pi_{ij}}\right) = \textbf{x}_{ij}^{\text{T}}\boldsymbol{\beta}_z + \textbf{d}_{j}^{\text{T}}\boldsymbol{\gamma}_z^{\left(g\right)} + \textbf{d}_{i}^{\text{T}}\boldsymbol{\gamma}_z^{\left(r\right)} + \theta_{zj}^{\left(g\right)} + \theta_{zi}^{\left(r\right)} + \nu_{zi}\nu_{zj}\end{equation} where $\textbf{x}_{ij}$ and $\textbf{d}_i$ were previously described in Section 3.1.  Because we now observe two responses for each pair (i.e., $T_{ij}$ and $T_{ji}$), we are able to separate the included parameters into the different roles of the individuals within a pair; specifically, we can estimate the infector or ``giver'' ($g$) and the infectee or ``receiver'' ($r$) terms.  For example, $T_{ij}$ describes the probability that individual $j$ transmits to $i$, so $\textbf{d}_j^{\text{T}} \boldsymbol{\gamma}_z^{(g)}$ from (4) represents the impact of the giver's (individual $j$'s) covariates on the probability of transmission while $\textbf{d}_i^{\text{T}} \boldsymbol{\gamma}_z^{(r)}$ describes the impact of the receiver's (individual $i$'s) characteristics.  Similarly, now each individual has two different additive random effect parameters, $\boldsymbol{\theta}_{zi}^{\text{T}} = \left(\theta_{zi}^{(g)}, \theta_{zi}^{(r)}\right)$; one for the giver and receiver roles, respectively.  The $\nu_{zi}\nu_{zj}$ interaction term accounts for correlation between observations from the same pair of individuals in different roles (i.e., $\nu_{zi}\nu_{zj} = \nu_{zj}\nu_{zi}$), where $\nu_{zi}|\sigma^2_{\nu_z} \stackrel{\text{iid}}{\sim}\text{N}\left(0, \sigma^2_{\nu_z}\right)$, $i=1,\hdots,n$ \citep{hoff2005bilinear}.  

In order to understand if the magnitude of a non-zero transmission probability is also impacted by covariates and random effect parameters, the pdf in (3) includes a separate regression framework for the non-zero probabilities.  Specifically, $f_{w_{ij}}\left(w\right)$ is a pdf introduced to model the non-zero probabilities on the logit scale such that \begin{equation} f_{w_{ij}}\left(w\right) \equiv \text{N}\left(\textbf{x}_{ij}^{\text{T}}\boldsymbol{\beta}_w + \textbf{d}_{j}^{\text{T}}\boldsymbol{\gamma}_w^{\left(g\right)} + \textbf{d}_{i}^{\text{T}}\boldsymbol{\gamma}_w^{\left(r\right)} + \theta_{wj}^{\left(g\right)} + \theta_{wi}^{\left(r\right)} + \nu_{wi}\nu_{wj},\ \sigma^2_{\epsilon}\right)\end{equation} where $\sigma^2_{\epsilon}$ represents the variance of the distribution and the remaining terms in (5) have been previously described.  The ``$w$'' subscripts in (5) serve to differentiate these parameters from those used by the binary model in (4) (i.e., ``$z$'' subscripts).

In total, each individual in the analysis has four additive random effect parameters, $\boldsymbol{\theta}_i^{\text{T}} = \left(\boldsymbol{\theta}_{zi}^{\text{T}}, \boldsymbol{\theta}_{wi}^{\text{T}}\right)$, representing the residual (i.e., after adjustment for known risk factors and repeated pair correlation) likelihood of transmitting ($g$) or being transmitted to ($r$) in the binary ($z$) and positive ($w$) transmission probability regressions.  Large values of the $z$ subscript random effect parameters suggest an increased chance of a non-zero transmission probability, while large values of the $w$ subscript parameters indicate an increasingly positive transmission probability.

As in (2), the introduced random effect parameters serve to adjust for pair- and proximity-based correlation in the outcomes.  Unlike in \cite{simpson2015two}, we anticipate that the collection of parameters corresponding to the same individual may themselves be correlated and introduce a multivariate model as a result.  The model for one set of the parameters is similar to (2) and is given as (for $i=1,\hdots,n)$ \begin{align*} \theta_{zi}^{(g)} = \eta_z^{(g)}\left\{h\left(\textbf{s}_i\right)\right\} + \zeta_{zi}^{(g)} \end{align*} where $\zeta_{zi}^{(g)}|\sigma^2_{\zeta^{(g)}_{z}} \stackrel{\text{iid}}{\sim}\text{N}\left(0, \sigma^2_{\zeta^{(g)}_{z}}\right)$ once again account for individual variability, and the remaining parameters, $\theta^{(r)}_{zi}$, $\theta^{(g)}_{wi}$, $\theta^{(r)}_{wi}$, are defined similarly.  To account for cross-covariance and spatial correlation among the parameters, we specify a multivariate Gaussian process for the mean parameters such that \begin{align}\begin{split} &\boldsymbol{\eta}|\Omega, \phi \sim \text{MVN}\left\{\boldsymbol{0}_{4m}, \Sigma\left(\phi\right) \otimes \Omega \right\} \text{ where}\\ &\boldsymbol{\eta}^{\text{T}} = \left\{\boldsymbol{\eta}\left(\textbf{s}^*_1\right)^{\text{T}}, \hdots, \boldsymbol{\eta}\left(\textbf{s}^*_m\right)^{\text{T}}\right\} \text{ and}\\ &\boldsymbol{\eta}\left(\textbf{s}^*_i\right)^{\text{T}} = \left\{\eta_z^{(g)}\left(\textbf{s}^*_i\right), \eta_z^{(r)}\left(\textbf{s}^*_i\right), \eta_w^{(g)}\left(\textbf{s}^*_i\right), \eta_w^{(r)}\left(\textbf{s}^*_i\right)\right\}.\end{split} \end{align}  The complete collection of mean parameters across all $m$ unique spatial locations is denoted by $\boldsymbol{\eta}$ (similar to (2)); $\boldsymbol{\eta}\left(\textbf{s}^*_i\right)$ represents the collection of mean parameters across each component of the model and different roles, specific to unique location $\textbf{s}_i^*$; $\Sigma\left(\phi\right)$ was previously described in (2); $\otimes$ is the Kronecker product; and $\Omega$ represents a four-by-four unstructured covariance matrix describing the cross-covariance among the set of four random effect parameters specific to a unique spatial location.
    
\subsubsection{Prior distributions}
We complete the model specification by assigning prior distributions to the introduced model parameters.  As in Section 3.1.1, the regression parameters are specified as $\beta_{zj}$, $\beta_{wj}$, $\gamma_{zk}^{(g)}$, $\gamma_{zk}^{(r)}$, $\gamma_{wk}^{(g)}$, $\gamma_{wk}^{(r)}$  $\stackrel{\text{iid}}{\sim}\text{N}\left(0,100^2\right)$ for $j=1,\hdots,p_x$ and $k=1,\hdots,p_d$; the variance parameters as $\sigma^2_{\epsilon}$, $\sigma^2_{\zeta_z^{(g)}}$, $\sigma^2_{\zeta_z^{(r)}}$, $\sigma^2_{\zeta_w^{(g)}}$, $\sigma^2_{\zeta_w^{(r)}}$, $\sigma^2_{\nu_z}$, $\sigma^2_{\nu_w}$ $\stackrel{\text{iid}}{\sim} \text{Inverse Gamma}\left(0.01, 0.01\right)$; the spatial correlation parameter as $\phi \sim \text{Gamma}\left(1.00, 1.00\right)$; and the cross-covariance matrix as $\Omega^{-1} \sim \text{Wishart}\left(5, I_4\right)$, a weakly informative choice resulting in uniform cross-correlations \textit{a priori} \citep{gelman2013bayesian}.

\subsection{Induced correlation structure}
The inclusion of individual-specific and spatially correlated random effect parameters in the dyadic outcome models detailed in Sections 3.1 and 3.2 results in a positive correlation between the responses whose magnitude varies depending on (i) if the two pairs share a common individual and (ii) the geographic distance between the people in the pairs.  To better understand the induced correlations, we consider two different cases specifically for patristic distances; when there is, and is not, a shared individual between the pairs.  Derivations for the transmission probability model are similar but more complicated due to the mixed-type distribution and two regression frameworks used.  Therefore, in Figure S1 of the Supplementary Material, we present simulation-based correlation estimates for this model over a range of spatial variance/correlation settings; see Section S1 for full details.   

First, we derive the correlation between two dyadic patristic distances that consist of entirely different individuals, $\ln\left(P_{ij}\right)$ and $\ln\left(P_{kl}\right)$ where $i \neq j \neq k \neq l$.  The variance of one of the responses (assuming every individual in the study has a unique spatial location) is given as \begin{align*}\begin{split} & \text{Var}\left\{\ln\left(P_{ij}\right)\right\} = \text{Var}\left(\theta_i\right) + \text{Var}\left(\theta_j\right) + \text{Var}\left(\epsilon_{ij}\right) + 2\text{Cov}\left(\theta_i,\theta_j\right)\\
&= \text{Var}\left\{\eta\left(\textbf{s}^*_i\right) + \zeta_i\right\} + \text{Var}\left\{\eta\left(\textbf{s}^*_j\right) + \zeta_j\right\} + \text{Var}\left(\epsilon_{ij}\right) + 2\text{Cov}\left\{\eta\left(\textbf{s}^*_i\right) + \zeta_i, \eta\left(\textbf{s}^*_j\right) + \zeta_j\right\}\\
&=2\tau^2\left(1 + \exp\left\{-\phi\left\|\textbf{s}^*_i - \textbf{s}^*_j \right\|\right\}\right) + 2\sigma^2_{\zeta} + \sigma^2_{\epsilon},\end{split}\end{align*} due to the spatial dependence between the $\theta_i$ parameters.  The covariance between these responses is given as \begin{align*} \begin{split} &\text{Cov}\left\{\ln\left(P_{ij}\right), \ln\left(P_{kl}\right)\right\} = \text{E}\left\{\left(\theta_i + \theta_j\right)\left(\theta_k + \theta_l\right)\right\} \\
&= \text{E}\left[\left\{\eta\left(\textbf{s}^*_i\right) + \zeta_i + \eta\left(\textbf{s}^*_j\right) + \zeta_j\right\}\left\{\eta\left(\textbf{s}^*_k\right) + \zeta_k + \eta\left(\textbf{s}^*_l\right) + \zeta_l\right\}\right] \\
&= \tau^2 \sum_{p_1 \in \left\{i,j\right\}} \sum_{p_2 \in \left\{k,l\right\}} \exp\left\{-\phi\left\|\textbf{s}^*_{p_1} - \textbf{s}^*_{p_2} \right\|\right\}, \end{split}\end{align*} and is simply a function of spatial distances between the individuals in the pairs.  This suggests that if spatial correlation is negligible then the covariance/correlation between outcome pairs without a shared individual is effectively equal to zero since $\exp\left\{-\phi\left\|\textbf{s}^*_i - \textbf{s}^*_j\right\|\right\} \approx 0$ for all $i,j$.  In the case of strong spatial correlation (i.e, $\exp\left\{-\phi\left\|\textbf{s}^*_i - \textbf{s}^*_j\right\|\right\} \approx 1$), the variance and covariance become $4\tau^2 + 2\sigma^2_{\zeta} + \sigma^2_{\epsilon}$ and $4\tau^2$, respectively.  This yields a correlation between the dyadic outcomes of $0$ and $\frac{4\tau^2}{4\tau^2 + 2\sigma^2_{\zeta} + \sigma^2_{\epsilon}}$ for the weak and strong spatial correlation settings, respectively.

Following similar derivations, the covariance between observations from two pairs that include one of the same individuals, say $\ln\left(P_{ij}\right)$ and $\ln\left(P_{ik}\right)$, is given as $$\tau^2\left(1 + \exp\left\{-\phi\left\|\textbf{s}^*_{i} - \textbf{s}^*_{j} \right\|\right\} +  \exp\left\{-\phi\left\|\textbf{s}^*_{i} - \textbf{s}^*_{k} \right\|\right\} + \exp\left\{-\phi\left\|\textbf{s}^*_{j} - \textbf{s}^*_{k} \right\|\right\}\right) + \sigma^2_{\zeta}.$$  Under negligible spatial correlation, this expression approaches $\tau^2 + \sigma^2_{\zeta}$ due to the fact that the same individual is represented in both pairs; while for strong spatial dependency, it approaches $4\tau^2 + \sigma^2_{\zeta}$.  This yields correlations of $\frac{\tau^2 + \sigma^2_{\zeta}}{2\tau^2 + 2\sigma^2_{\zeta} + \sigma^2_{\epsilon}}$ and $\frac{4\tau^2 + \sigma^2_{\zeta}}{4\tau^2 + 2\sigma^2_{\zeta} + \sigma^2_{\epsilon}}$ for the weak and strong spatial correlation settings, respectively.  Both of these quantities are larger than the comparable versions for pairs that do not share an individual.

These findings, along with the results shown in Figure S1 of the Supplementary Material, suggest that there is increased correlation between outcomes when at least one individual is shared between the pairs.  In the transmission probability model, there is also extremely high correlation between outcomes when the pairs include both individuals but in different roles, with spatial correlation/variability having almost no impact.  However, even when pairs have no shared individuals, strong correlation can still exist depending on the strength/magnitude of spatial correlation/variability in the data.  This may be an important feature to consider in the infectious disease setting where unmeasured transmission dynamics could results in residual spatial correlation between individuals.

\section{Simulation study}
We design a simulation study to investigate the implications of ignoring multiple forms of correlation between dyadic genetic relatedness outcomes when making inference on regression parameters of interest.  Additionally, we aim to determine if a common Bayesian model comparison tool can be used to identify datasets that contain non-negligible levels of correlation and also differentiate the sources of the correlation.  In order to avoid repetition of text, the process described throughout Section 4 is specifically for the patristic distances model.  However, we carried out the same steps for transmission probabilities using the corresponding equations from Section 3.2.  These full details are provided in Section S2 of the Supplementary Material.

\subsection{Data generation}
We simulate data from the model in (1) under three different scenarios.  In Setting 1, we specify that there is no unmeasured correlation by setting $\theta_i = 0$ for all $i$.  In Setting 2, we simulate data from (1) and (2) with $\eta\left(\textbf{s}_i^*\right) = 0$ for all $i$, resulting in only non-spatial variability in the random effect parameters.  In Setting 3, we also simulate data from (1) and (2) with $\phi = -\ln\left(0.05\right)/\max\left\{\left\|\textbf{s}_i^* - \textbf{s}_j^*\right\|; i < j\right\}$ and $\zeta_i = 0$ for all $i$, resulting in spatially correlated random effect parameters with correlation that decreases to $0.05$ at the maximum distance observed in the data (i.e., the effective range). 

When simulating data from these models, we use results from our data application in the Republic of Moldova (Section 5) to ensure that we are working with realistic outcomes.  Specifically, we use the same sample size ($n=99$), same covariates ($\textbf{x}_{ij}$, $\textbf{d}_i$), and choose the true parameter values needed to simulate from (1) and (2) based on posterior estimates obtained from the data application.  In Table S1 of the Supplementary Material, the specific values used in each simulation study are given.  

When simulating data from Settings 2 and 3, we define the total variance of the random effect process as $\tau^2 + \sigma^2_{\zeta}$ and use the values in Table S1 of the Supplementary Material to estimate this quantity.  In Setting 2, all of this variability is assumed to be non-spatial while in Setting 3 it is attributed entirely to spatial correlation.  We generate a new vector of $\boldsymbol{\theta}$ parameters for each dataset from the setting-specific model.  We also create a unique set of spatial locations, with no co-located individuals (i.e., $m = n$), for each dataset.  In total, we simulate 100 datasets from each setting.

\subsection{Competing models}
We apply three competing models to every dataset.  The first model (i.e., \textit{Fixed}) represents a simplified fixed effects regression form of (1) where $\theta_i = 0$ for all $i$ (i.e., no random effect parameters).  As \textit{Fixed} matches the data generated from Setting 1, we expect it to perform well overall in that setting.  However, in Settings 2 and 3, it will likely struggle in estimating the associations of interest as it ignores all correlation.  The second model (i.e., \textit{Non-spatial}) also represents a variant of (1) and (2) where $\eta\left(\textbf{s}^*_i\right) = 0$ for all $i$ (matching data generation Setting 2).  Therefore, \textit{Non-spatial} accounts for correlation due to the nature of dyadic responses, but ignores the potential for spatial correlation in the data.  It is currently unknown how inference for the regression parameters is impacted when dyadic correlation is accounted for but spatial correlation is ignored.  The final competing model (i.e., \textit{Spatial}) is our newly developed model in Section 3.1 which can address both sources of correlation simultaneously.    

We monitor several pieces of information collected from the analyzed datasets to compare the methods.  First, we calculate the mean absolute error (MAE) for every regression parameter in the model using the posterior mean as the point estimate.  Next, we calculate 95\% quantile-based equal-tailed credible intervals (CIs) for each regression parameter and monitor how often this interval includes the true value (ideally around 95\% of the time) and its length.  Finally, we formally compare the models using Watanabe Akaike information criterion (WAIC), a metric that balances model fit and complexity where smaller values suggest that a model is preferred \citep{watanabe2010asymptotic}.

\subsection{Results}
We apply each method to each dataset and collect 20,000 posterior samples after removing the first 5,000 iterations prior to convergence of the model.  Additionally, we thin the remaining samples by a factor of two to reduce posterior autocorrelation, resulting in 10,000 samples with which to make posterior inference.  The priors from Sections 3.1.1 and 3.2.1 were used other than when we encountered convergence issues when applying \textit{Spatial} in the transmission probabilities framework to a few of the datasets generated from Setting 3.  In that case, $\sigma^2_{\nu_z} \sim \text{Inverse Gamma}\left(100.00, 100.00\right)$ or $\text{Inverse Gamma}\left(1000.00, 1000.00\right)$ was used to stabilize estimation of the $\nu_{zi}$ parameters.  

The full results are shown in Table 2 for all models where we report the average MAE, average empirical coverage (EC), and average CI length across all regression parameters and simulated datasets, as well as the average difference in WAIC values, with respect to \textit{Spatial}, across all simulated datasets. 

\begin{landscape}
\begin{table}
\centering
\caption{Simulation study results.  $\Delta$ WAIC = WAIC $-$ WAIC$_{\text{new}}$.  Averages across the 100 simulated datasets are reported with standard errors given in parentheses.  Bold entries indicate the ``best'' value within a setting and outcome type across models.  MAE results for the patristic distances model are multiplied by 100 for presentation purposes.}
\begin{tabular}{llrrrrrrr}
\hline
 & &  \multicolumn{3}{c}{Patristic} & & \multicolumn{3}{c}{Trans.\ Probs.} \\
 \cline{3-5} \cline{7-9}
Metric        & Setting & Standard             & Non-spatial  & Spatial              & & Standard         & Non-spatial    & Spatial              \\
\hline
MAE           &  1      & \textbf{1.02} (0.03) & 1.03 (0.03)  & 1.04 (0.03)          & & 1.62 (0.20)      & 1.53 (0.19)    & \textbf{1.40} (0.16) \\
              &  2      & 6.77 (0.23)          & 6.21 (0.23)  & \textbf{6.19} (0.23) & & 0.40 (0.00)      & 0.35 (0.01)    & \textbf{0.34} (0.01) \\
              &  3      & 5.56 (0.20)          & 4.84 (0.19)  & \textbf{2.31} (0.08) & & 0.43 (0.03)      & 0.61 (0.24)    & \textbf{0.35} (0.08) \\
\hline
EC            &  1      & \textbf{0.94} (0.01) & 0.97 (0.00)  & 0.97 (0.00)          & & 0.92 (0.01)      & 0.92 (0.01)    & \textbf{0.93} (0.01) \\
              &  2      & 0.41 (0.02)          & 0.93 (0.01)  & \textbf{0.94} (0.01) & & 0.44 (0.01)      & 0.93 (0.00)    & \textbf{0.94} (0.00) \\
              &  3      & 0.39 (0.01)          & 0.92 (0.01)  & \textbf{0.96} (0.01) & & 0.47 (0.01)      & 0.90 (0.01)    & \textbf{0.94} (0.01) \\
\hline
CI Length     &  1      & 0.05 (0.00)          & 0.06 (0.00)  & 0.06 (0.00)          & & 4.33 (0.41)      & 4.26 (0.40)    & 4.28 (0.39)          \\
              &  2      & 0.10 (0.00)          & 0.28 (0.00)  & 0.30 (0.00)          & & 0.38 (0.00)      & 1.61 (0.01)    & 1.62 (0.01)          \\
              &  3      & 0.08 (0.00)          & 0.21 (0.00)  & 0.13 (0.00)          & & 0.52 (0.09)      & 1.90 (0.48)    & 1.31 (0.18)          \\
\hline
$\Delta$ WAIC &  1      & -31.74 (0.97)        & -2.95 (0.07) & --                   & & 11.28 (1.95)     & -0.49 (1.74)   & --                   \\
              &  2      & 6138.68 (56.90)      & 0.57 (0.02)  & --                   & & 9013.90 (87.72)  & 26.04 (22.47)  & --                   \\
              &  3      & 3983.08 (98.48)      & 4.89 (0.10)  & --                   & & 8349.95 (298.65) & 111.87 (10.67) & --                   \\
\hline
\end{tabular}
\end{table}
\end{landscape}
\clearpage

As expected, all competing models perform similarly in terms of MAE, EC, and CI length in Setting 1, with \textit{Spatial} having improved MAE results for the transmission probabilities framework.  For the patristic distances model, WAIC also correctly favors \textit{Fixed} on average as it most closely matches the way the data were generated.  However, with respect to WAIC, \textit{Fixed} is slightly outperformed by \textit{Spatial} in this setting for transmission probabilities.  

In Settings 2 and 3 where correlation is present, \textit{Spatial} consistently outperforms \textit{Fixed} and \textit{Non-spatial} across all metrics.  \textit{Fixed} displays troubling behavior in these settings, particularly with respect to EC.  The 95\% CIs are only capturing the true parameter values around 39-47\% of the times and are much shorter on average than those from the competing methods.  This suggests that failing to account for correlation may result in CIs that are too narrow, possibly leading to an inflated type I error rate.  

The MAE results suggest that regression parameter inference may also suffer when correlation is ignored and/or mischaracterized.  For example, in Setting 3 where the data exhibit spatial correlation, \textit{Spatial} produces substantially smaller MAEs than \textit{Non-spatial} (and \textit{Fixed}), showing the importance of accounting for spatial correlation.  However, in Setting 2 where there is no spatial correlation, \textit{Non-spatial} and \textit{Spatial} perform almost identically with respect to MAE.  

The WAIC results in Settings 2 and 3 are decisively in favor of \textit{Spatial} over \textit{Fixed}, with differences between \textit{Spatial} and \textit{Non-spatial} increasing under spatially correlated data.  When combined with results from Setting 1, this provides evidence to suggest that WAIC may be a useful tool for differentiating datasets in terms of the presence and composition of correlation.  

\section{\emph{Mycobacterium tuberculosis} in the Republic of Moldova}
We apply the methods developed in Sections 3.1 and 3.2 to better understand factors related to \emph{Mycobacterium tuberculosis} transmission dynamics in the Republic of Moldova.  Specifically, $\textbf{x}_{ij}$ from (1, 4, 5) includes the pair-specific covariates described in Section 2 and Table 1 (i.e., intercept, same village indicator, distance between villages, difference in diagnosis dates, difference in ages) while $\textbf{d}_i$ includes the individual-specific covariates (i.e., age, sex, education, working status, residence type).  In addition to the new methods, we also present results from the competing approaches detailed in Section 4.2 (i.e., \textit{Fixed} and \textit{Non-spatial}).  We use WAIC to compare the different models, identify the source of correlation, and ultimately determine the need for the new methodology.

All models are fit in the Bayesian setting using MCMC sampling techniques, with the full conditional distributions detailed in Section S3 of the Supplementary Material.  For each method, we collect 10,000 samples from the joint posterior distribution after removing the first 50,000 iterations prior to convergence and thinning the remaining 200,000 posterior samples by a factor of 20 to reduce posterior autocorrelation.  For each parameter, convergence was assessed using Geweke's diagnostic \citep{geweke1991evaluating} while effective sample size was calculated to ensure we collected sufficient post-convergence samples to make accurate statistical inference.  Neither tool suggested any issues of concern.  We present posterior means and 95\% quantile-based equal-tailed CIs when discussing posterior inference.  

\subsection{Patristic distances}
Results from the patristic distance analyses are shown in Figure 2 and Table 3.  From Figure 2 it is clear that WAIC favors \textit{Spatial} over the other approaches.  In the simulation study, WAIC was shown to consistently identify the correct data generating setting, suggesting that there is likely non-negligible \textbf{and} spatially structured correlation in these data.  Consequently, \textit{Fixed} should not be used as it is likely to underestimate regression parameter uncertainty, resulting in CIs that are often too narrow, while \textit{Non-spatial} may result in regression parameter estimates with inflated MAE.  This can partly be seen in Figure 2 where the \textit{Fixed} CIs are much shorter on average.  The point estimates seen in Figure 2 are generally consistent across all methods in this case.  

\begin{figure}[ht!]
\begin{center}
\includegraphics[scale = 0.45]{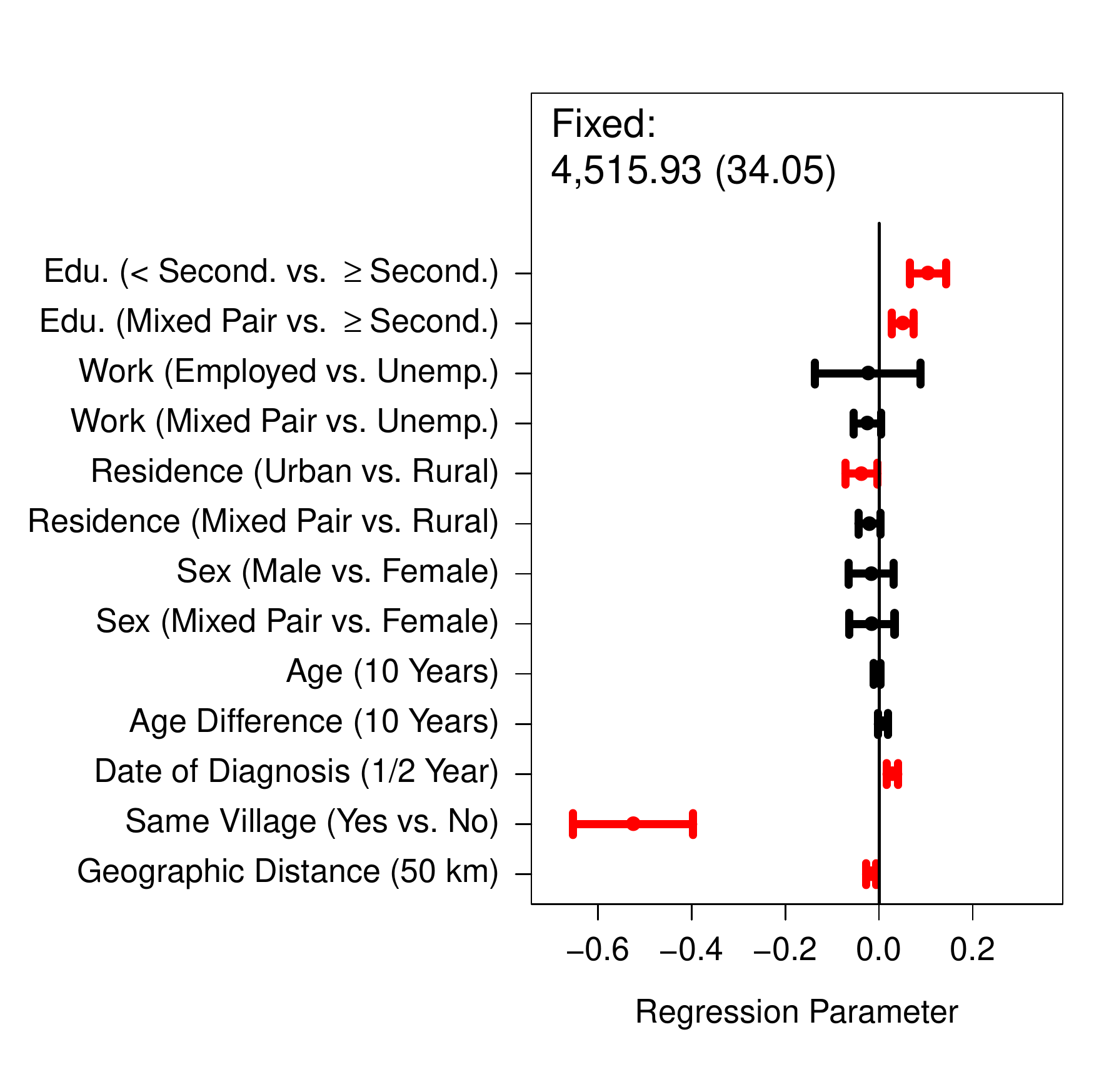}
\includegraphics[trim={8.6cm 0.0cm 0.0cm 0.0cm}, clip, scale = 0.45]{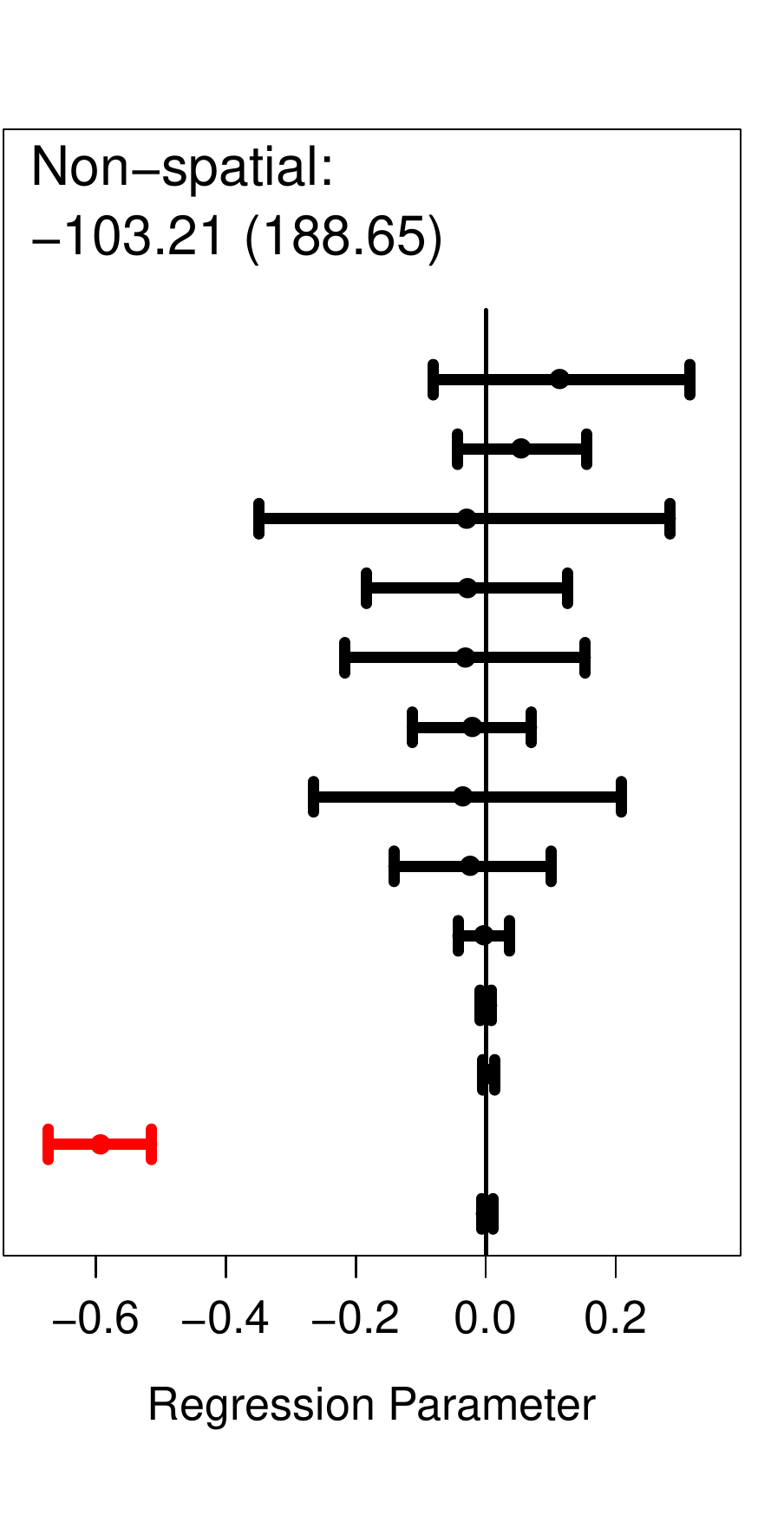}
\includegraphics[trim={8.6cm 0.0cm 0.0cm 0.0cm}, clip, scale = 0.45]{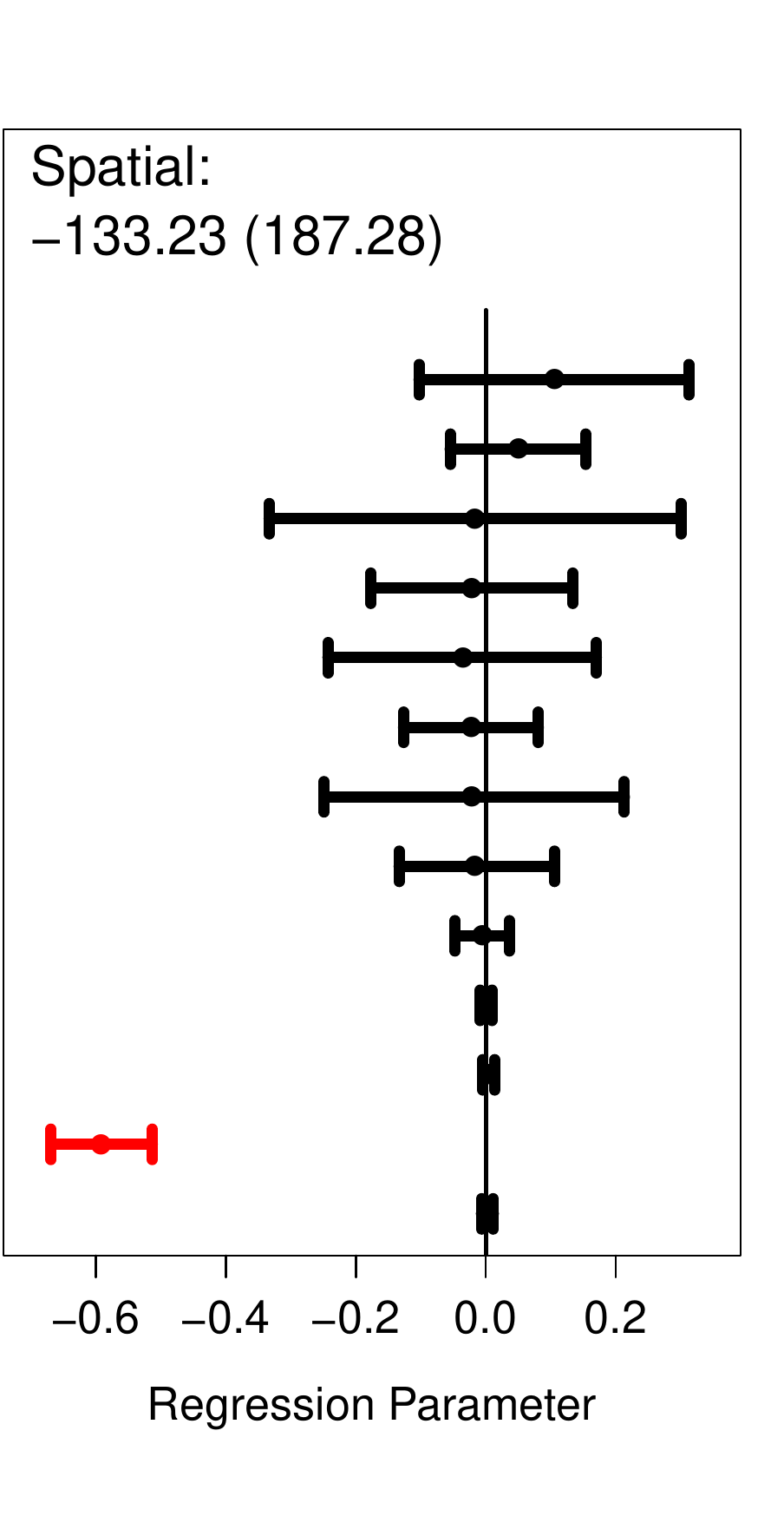}\\
\caption{Results from the patristic distance analyses in the Republic of Moldova.  Posterior means and 95\% quantile-based equal-tailed credible intervals are shown for the raw regression parameters from each of the competing methods.  Red lines indicate that the interval excludes zero.  WAIC values are provided along with the expected number of parameters/model complexity term in parentheses (i.e., p$_{\text{WAIC}}$), with smaller values of WAIC preferred.}
\end{center}
\end{figure}

Posterior inference from \textit{Spatial} for the exponentiated regression parameters in Table 3 suggest that the indicator of whether the pair of individuals was in the same village is the only association whose CI excludes one.  Patristic distances from pairs of individuals from the same village are around 45\% (40\%, 49\%) smaller on average than those from pairs of individuals in different villages. 

\begin{table}
\centering
\caption{Results from the \textit{Spatial} patristic distance analyses in the Republic of Moldova.  Posterior means and 95\% quantile-based equal-tailed credible intervals are shown for the exponentiated regression parameters (i.e., ratio of expected patristic distances per specified change in covariate value).  The displayed credible intervals for the bolded entries exclude one.}
\begin{tabular}{lr}
\hline
Effect                                              & Estimate (95\% CI)        \\
\hline
Distance Between Villages (50 km)                   & 1.00 (0.99, 1.01)          \\
Same Village (Yes vs.\ No)                          & \textbf{0.55 (0.51, 0.60)} \\
Date of Diagnosis Difference (1/2 Year)             & 1.01 (1.00, 1.01)          \\
Age Difference (10 Years)                           & 1.00 (0.99, 1.01)          \\
Age (10 Years)                                      & 1.00 (0.95, 1.04)          \\
Sex:                                                &                            \\
\ \ \ Mixed Pair vs.\ Both Female                   & 0.99 (0.88, 1.11)          \\
\ \ \ Both Male vs.\ Both Female                    & 0.99 (0.78, 1.24)          \\
Residence Location:                                 &                            \\
\ \ \ Mixed Pair vs.\ Both Rural                    & 0.98 (0.88, 1.08)          \\
\ \ \ Both Urban vs.\ Both Rural                    & 0.97 (0.78, 1.19)          \\
Working Status:                                     &                            \\
\ \ \ Mixed Pair vs.\ Both Unemployed               & 0.98 (0.84, 1.14)          \\
\ \ \ Both Employed vs.\ Both Unemployed            & 1.00 (0.72, 1.35)          \\
Education:                                          &                            \\
\ \ \ Mixed Pair vs.\ Both $\geq$ Secondary         & 1.05 (0.95, 1.17)          \\
\ \ \ Both $<$ Secondary vs.\ Both $\geq$ Secondary & 1.12 (0.90, 1.37)          \\
\hline
\end{tabular}
\end{table}

\subsection{Transmission probabilities}
Results from the transmission probability analyses are shown in Figure 3 and Table 4.  Similar to the patristic distance analyses, WAIC also favors \textit{Spatial} over the competing approaches in this setting.  Once again, \textit{Fixed} may be producing CIs that are too narrow while the \textit{Non-spatial} point estimates may have inflated MAEs.  Graphical results from \textit{Spatial} show several more significant associations than observed with the competing approaches.  Recall that in the transmission probabilities framework, \textit{Non-spatial} specifies that $\boldsymbol{\eta}\left(\textbf{s}^*_i\right) = \textbf{0}_4$ for all $i$, meaning that it accounts for neither spatial correlation nor cross-covariance between the parameters.  This could explain the relatively large differences observed between results from the two methods.  

\begin{figure}[ht!]
\centering
\includegraphics[trim={0.0cm 1.0cm 0.0cm 0.0cm}, clip, scale = 0.45]{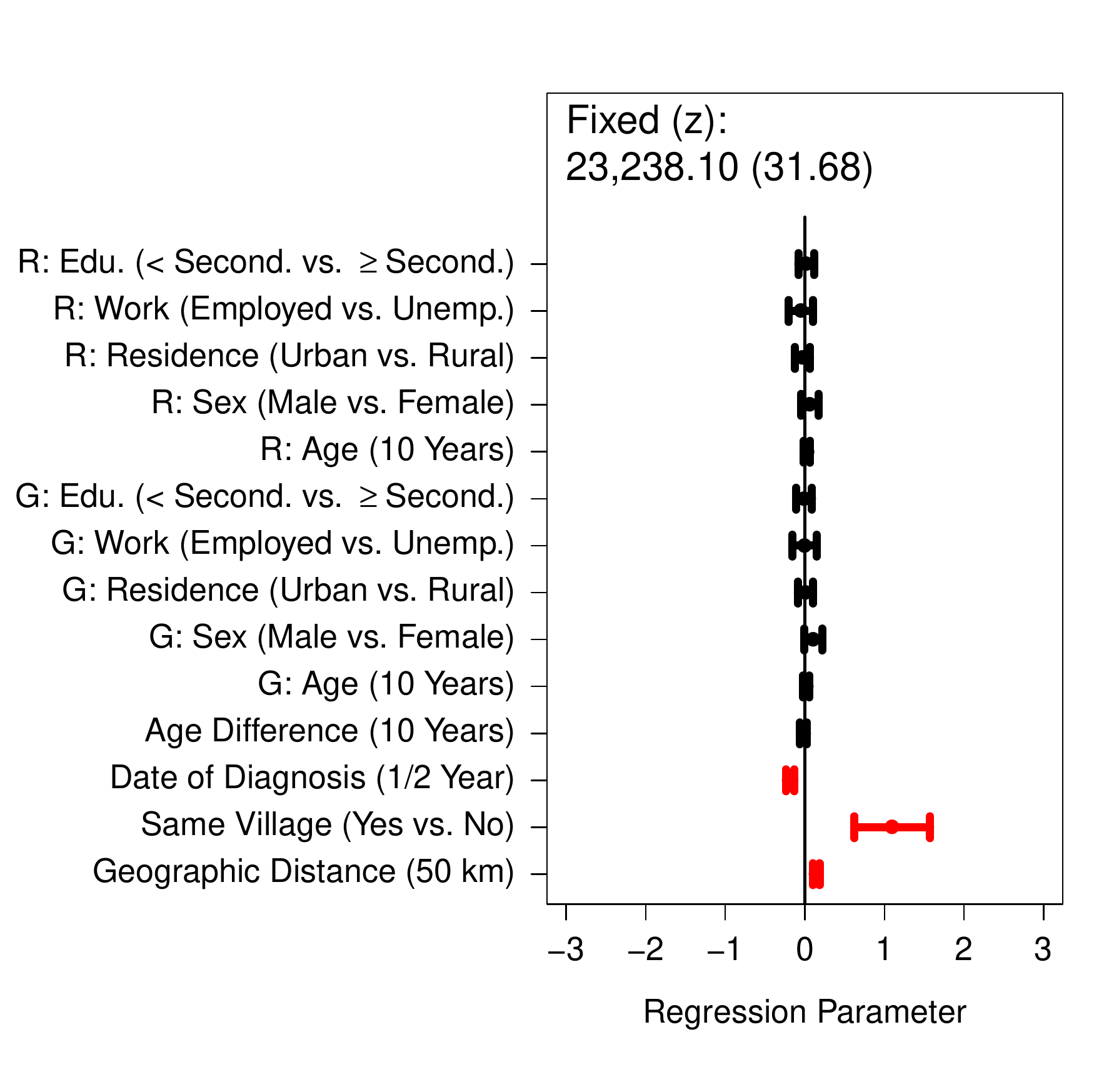}
\includegraphics[trim={8.6cm 1.0cm 0.0cm 0.0cm}, clip, scale = 0.45]{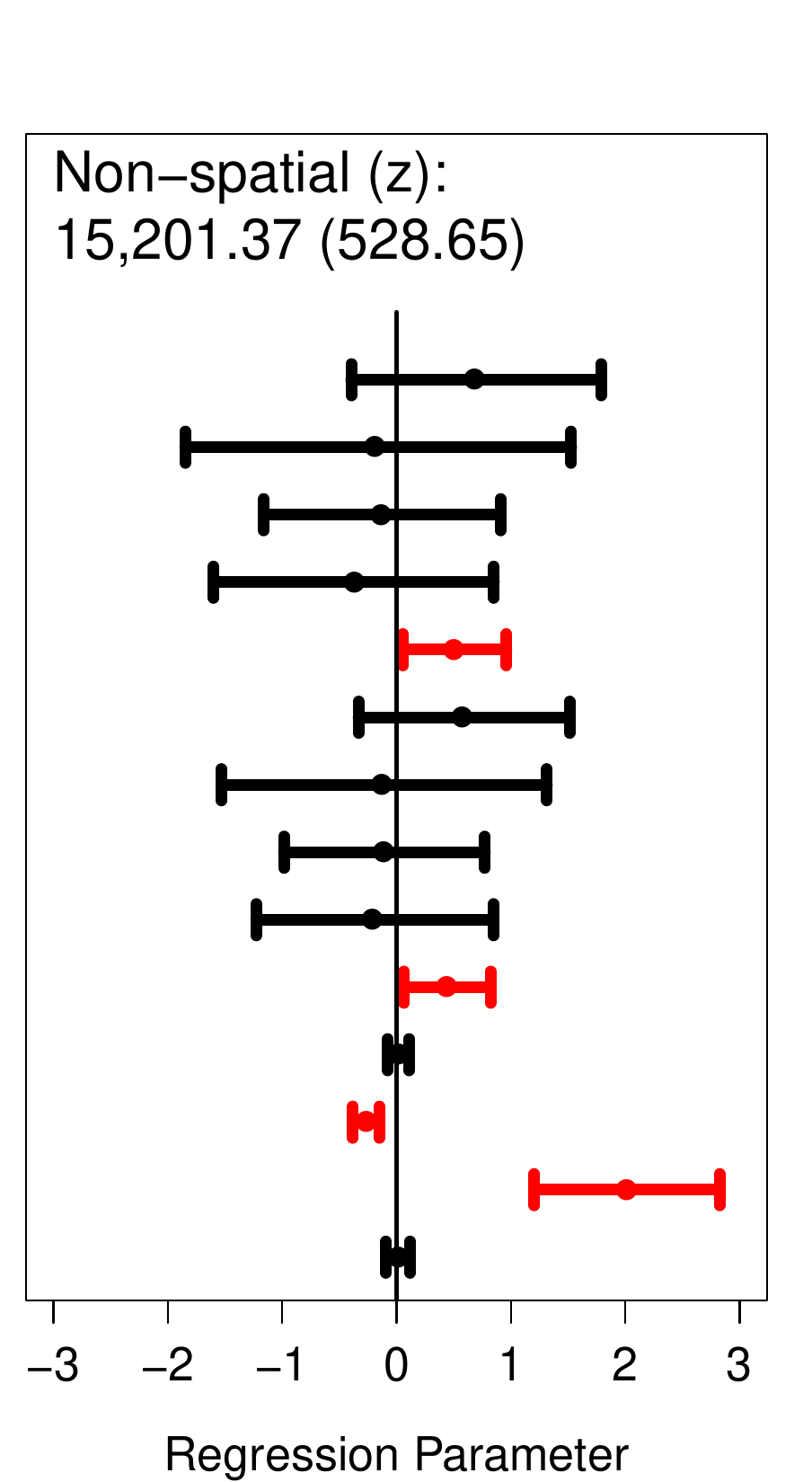}
\includegraphics[trim={8.6cm 1.0cm 0.0cm 0.0cm}, clip, scale = 0.45]{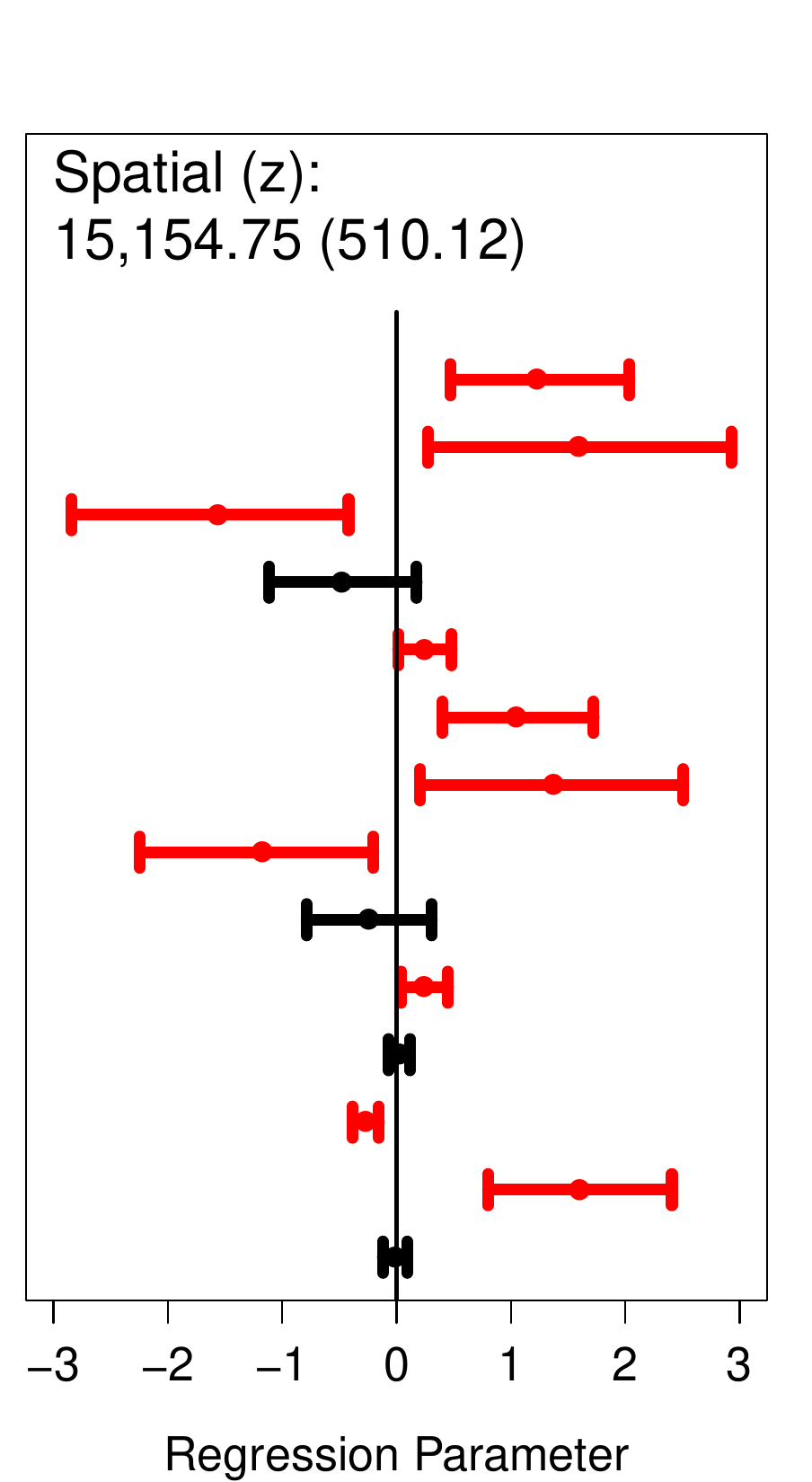}\\
\includegraphics[scale = 0.45]{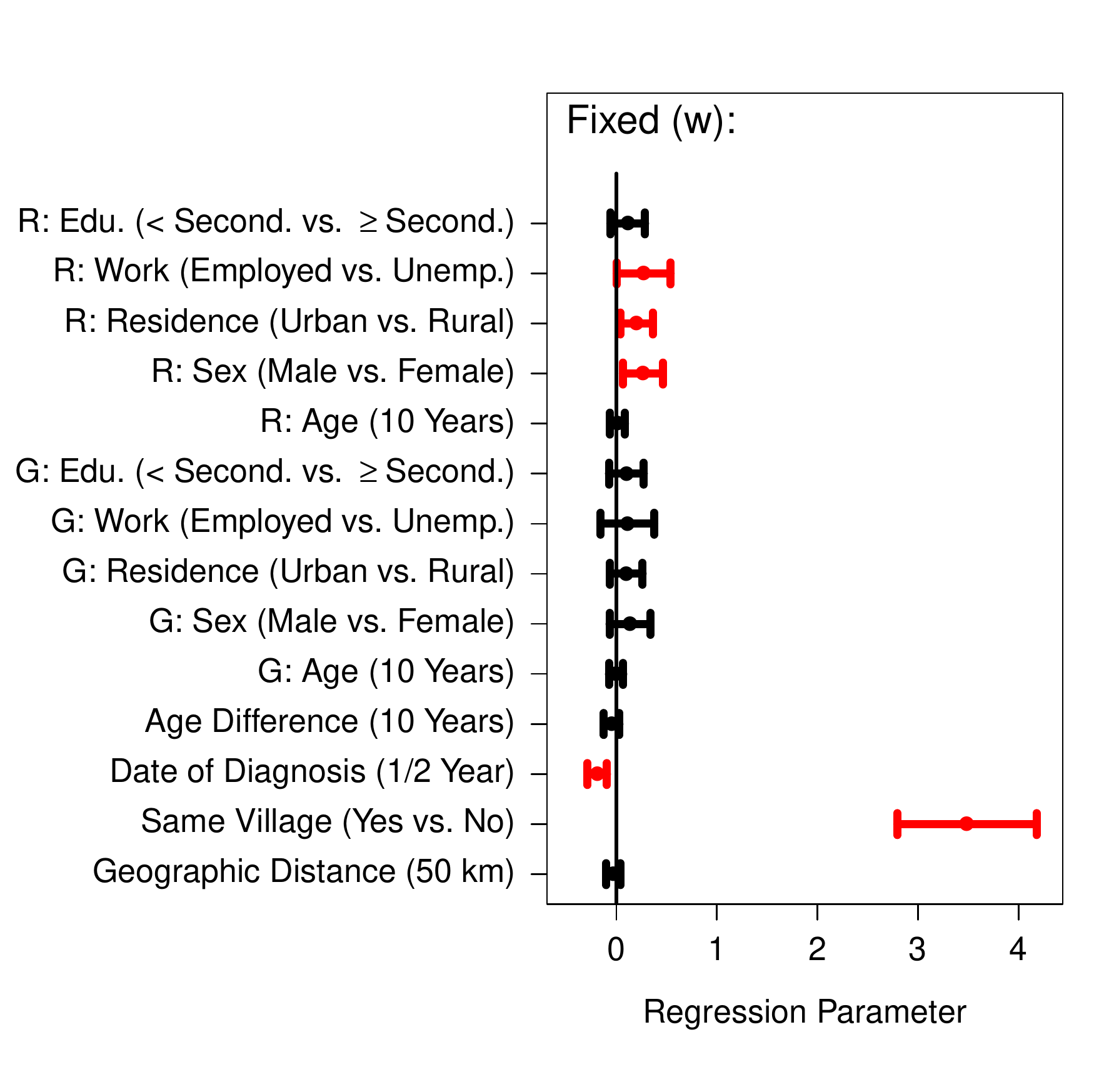}
\includegraphics[trim={8.6cm 0.0cm 0.0cm 0.0cm}, clip, scale = 0.45]{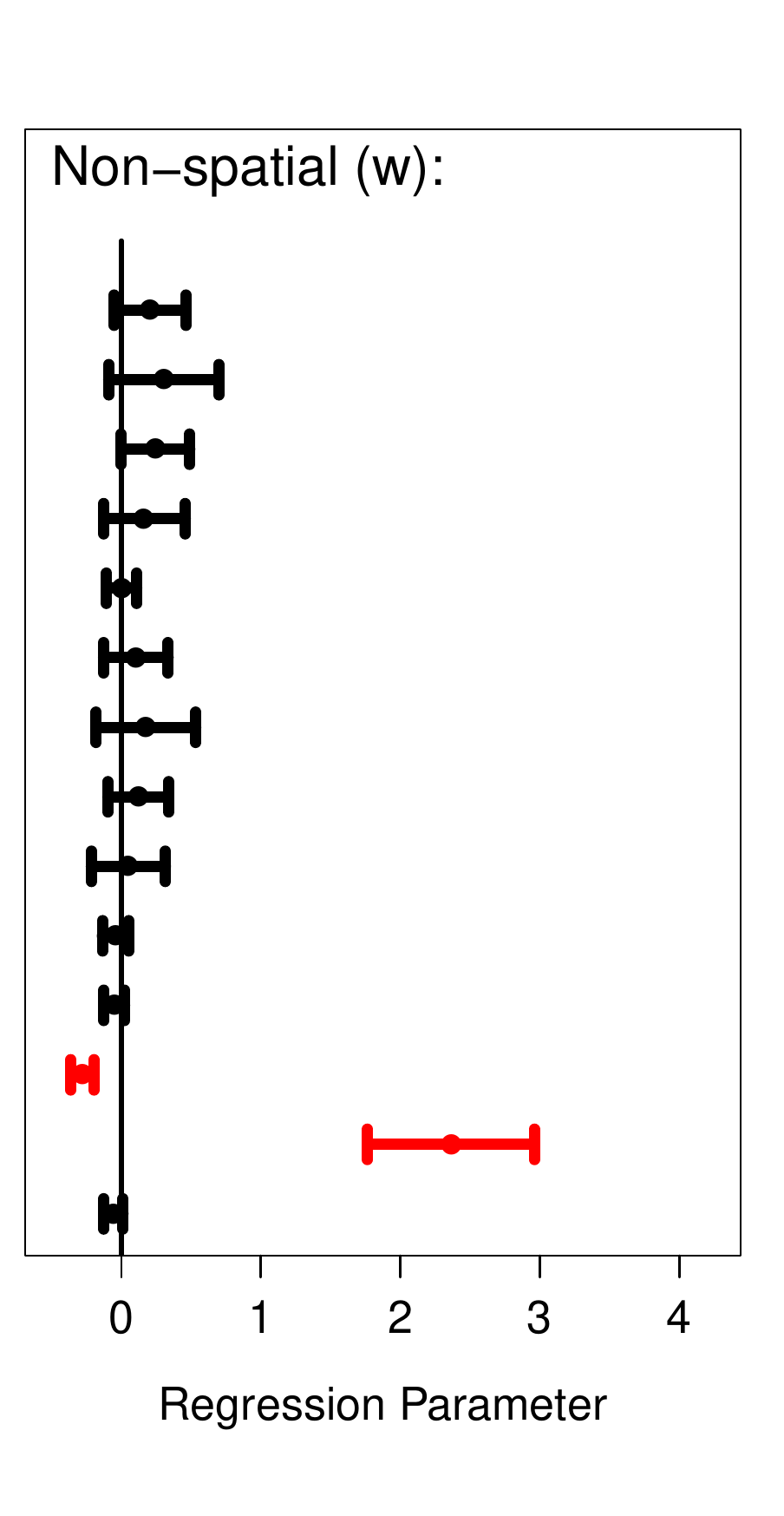}
\includegraphics[trim={8.6cm 0.0cm 0.0cm 0.0cm}, clip, scale = 0.45]{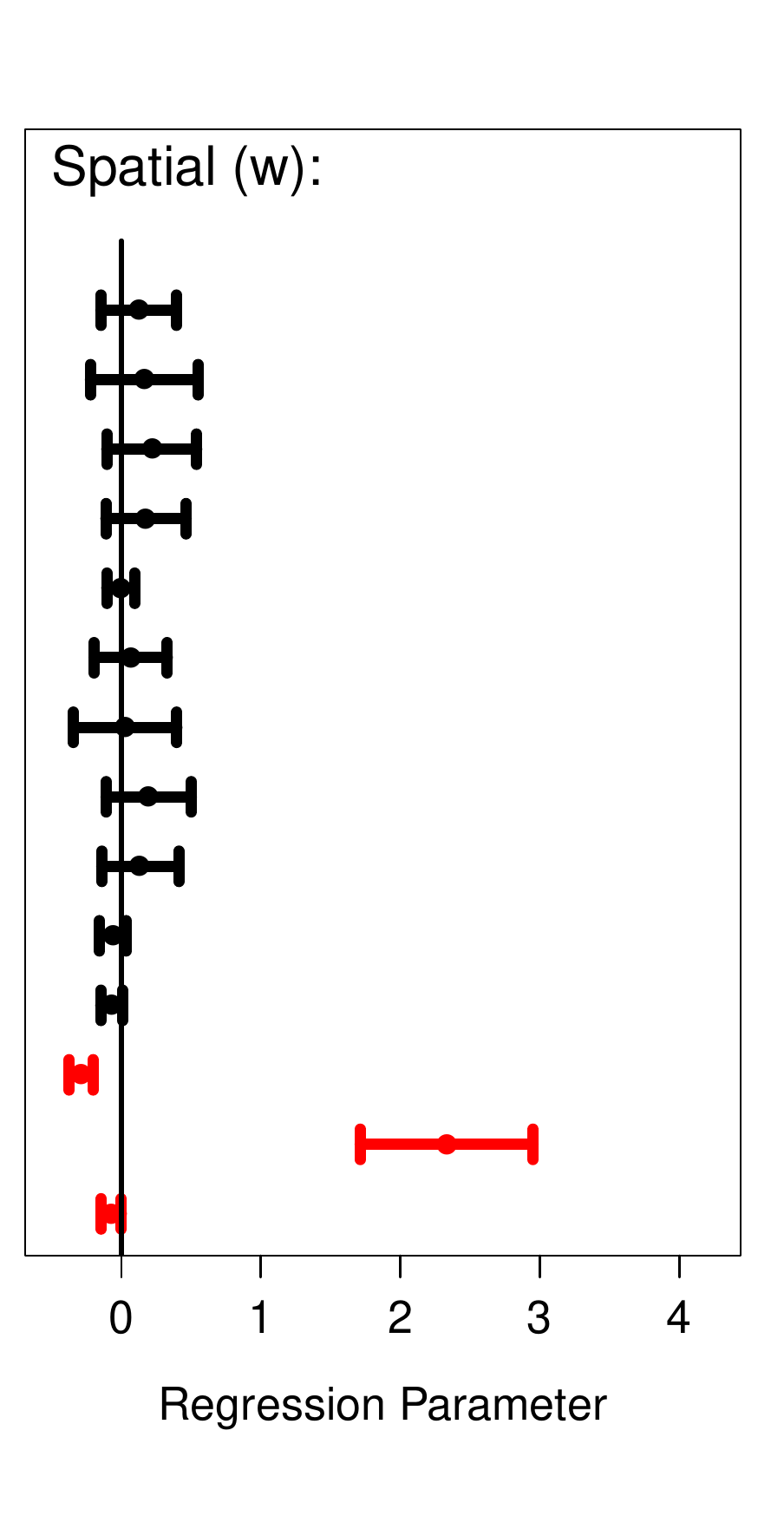}
\caption{Results from the transmission probability analyses in the Republic of Moldova.  Posterior means and 95\% equal-tailed quantile-based credible intervals are shown for the binary component (top row) and continuous component (bottom row) raw regression parameters from each of the competing methods.  Red lines indicate that the interval excludes zero.  WAIC values are provided along with the expected number of parameters/model complexity term in parentheses (i.e., p$_{\text{WAIC}}$), with smaller values of WAIC preferred.}
\end{figure}
\clearpage 

For the pair-specific covariates in the binary regression model, odds ratio results in Table 4 from \textit{Spatial} suggest that pairs of individuals in the same village and with similar dates of diagnosis are more likely to have non-zero transmission probabilities on average.  Results from the continuous regression model are similar, with the addition of a significant finding for the distance between villages.  Pairs of individuals whose villages are closer together geographically are more likely to have larger non-zero transmission probabilities. 

\begin{table}[ht!]
\small 
\centering
\caption{Results from the \textit{Spatial} transmission probability analyses in the Republic of Moldova.  Posterior means and 95\% quantile-based equal-tailed credible intervals are shown for the exponentiated regression parameters (i.e., odds ratios).  The displayed credible intervals for the bolded entries exclude one.}
\begin{tabular}{lrrr}
\hline
Effect                                     & Binary & &  Continuous           \\
\hline
Distance Between Villages (50 km)          & 0.99 (0.89, 1.10)          & & \textbf{0.93 (0.87, 1.00)} \\
Same Village (Yes vs.\ No)                 & \textbf{5.41 (2.22, 11.11)} & & \textbf{10.84 (5.54, 19.12)} \\
Date of Diagnosis Difference (1/2 Year)    & \textbf{0.77 (0.68, 0.86)} & & \textbf{0.75 (0.69, 0.82)} \\
Age Difference (10 Years)                  & 1.03 (0.93, 1.13)          & & 0.93 (0.86, 1.01)         \\
\hline 
\textbf{Giver Role:}                       &            & &                 \\
Age (10 Years)                             & \textbf{1.28 (1.04, 1.57)} &  & 0.94 (0.86, 1.03)        \\
Sex:                                       &                   &  &              \\
\ \ \ Male vs.\ Female                     & 0.82 (0.46, 1.36) &  & 1.15 (0.87, 1.51)        \\
Residence Location:                        &                   &  &              \\
\ \ \ Urban vs.\ Rural                     & \textbf{0.35 (0.11, 0.82)} &  & 1.23 (0.90, 1.65)        \\
Working Status:                            &                   &  &              \\
\ \ \ Employed vs.\ Unemployed             & \textbf{4.69 (1.23, 12.26)}&  & 1.05 (0.71, 1.49)        \\
Education:                                 &                   &  &              \\
\ \ \ $<$ Secondary vs.\ $\geq$ Secondary  & \textbf{3.01 (1.49, 5.61)} &  & 1.08 (0.82, 1.39)        \\
\hline
\textbf{Receiver Role:}                    &                   &  &                          \\
Age (10 Years)                             & \textbf{1.28 (1.01, 1.62)} &  & 1.00 (0.90, 1.10)        \\
Sex:                                       &                   &  &              \\
\ \ \ Male vs.\ Female                     & 0.65 (0.33, 1.19) &  & 1.20 (0.90, 1.59)        \\
Residence Location:                        &                   &  &              \\
\ \ \ Urban vs.\ Rural                     & \textbf{0.25 (0.06, 0.66)} &  & 1.27 (0.90, 1.72)        \\
Working Status:                            &                   &  &              \\
\ \ \ Employed vs.\ Unemployed             & \textbf{6.18 (1.32, 18.66)}&  & 1.20 (0.80, 1.73)        \\
Education:                                 &                   &  &              \\
\ \ \ $<$ Secondary vs.\ $\geq$ Secondary  & \textbf{3.70 (1.60, 7.66)} &  & 1.15 (0.86, 1.48)        \\
\hline 
\end{tabular}
\end{table}
\clearpage 

For the individual-specific covariates, overall the giver and receiver results are very similar within both the binary and continuous regression models, suggesting that the impact of the included factors do not differ based on what role the individual is serving in the pair.  For the binary component of the model we see that older, employed, and less formally educated individuals living in rural areas are more likely to be in pairs that have non-zero transmission probabilities.  There were no statistically significant individual-level associations identified for the continuous component of the model.     

\subsection{Random effect parameter analyses}
In Section S4 of the Supplementary Material, we present two additional analyses for the estimated random effect parameters, $\theta_i$ and $\left(\theta_{zi}^{(g)}, \theta_{zi}^{(r)}, \theta_{wi}^{(g)}, \theta_{wi}^{(r)}\right)^{\text{T}}$, from the fitted patristic distances and transmission probabilities models, respectively.  First, we present several summaries of the estimated random effect parameters/hyperparameters with respect to the observed dyadic genetic distance outcomes, finding an intuitive connection between the random effect parameters and outcomes as well as the clear presence of spatially structured variability in the data.  Results suggest that the frameworks have the ability to identify key individuals associated with increased transmission activity.  Next, we use Bayesian kriging techniques \citep{banerjee2014hierarchical} to predict the random effect parameters at new/unobserved spatial locations for both outcomes across the Republic of Moldova.  Posterior predicted mean and standard deviation maps are presented which allow us to determine areas of increased residual transmission risk for both outcomes.

\section{Discussion} 
In this work, we presented innovative hierarchical Bayesian spatial methods for modeling two types of dyadic genetic relatedness data, patristic distances and transmission probabilities.  The models account for multiple sources of correlation (i.e., dyadic and spatial) and important features of each outcome (e.g., zero-inflation).  Through simulation, we showed that these approaches perform as well as \textit{Fixed} (i.e., regression while ignoring correlation) and \textit{Non-spatial} (i.e., no spatial or cross-covariance) when applied to uncorrelated data, and greatly outperform them under correlation in terms of estimating and quantifying uncertainty in the regression parameters.  Under any type of correlation, \textit{Fixed} produces CIs that are too narrow and as a result should not be used for estimating associations between genetic relatedness measures and other factors.  

When applying the models to \emph{Mycobacterium tuberculosis} data from the Republic of Moldova, we found significant associations between genetic relatedness and spatial proximity, dates of diagnosis, and multiple individual-level factors; many of which represent new insights in this setting.  Analysis of the random effect parameters identified individuals and geographic areas that are generally associated with higher levels of transmission, a potentially useful aspect of the model with respect to infection control as heterogeneity in infectiousness among TB patients has been investigated in previous studies \citep{ypma2013sign, melsew2019role}.  Additionally, WAIC showed that the correlation between outcomes was at least partially spatially structured, further motivating the development of the new methodology.

The unique data in our study represent a major strength of the work, however, a number of factors inherent to \emph{Mycobacterium tuberculosis} make the study of transmission dynamics difficult.  First, TB epidemics are relatively slow moving, often occurring over years \citep{Pai}, and it is difficult to observe these processes in a two-year study.  Second, the time between infection and disease onset varies greatly, from weeks to years \citep{Borgdorff}.  Third, not all TB cases are `culture positive', meaning it is more difficult to culture, and, by extension, sequence isolates for every known TB case \citep{Cruciani, Nguyen}.  Taken together, these factors make it difficult to infer direct transmission events.  Even within a putative transmission cluster, the probabilities of direct transmission among case pairs may be low.  For this reason, it can be difficult to fit a model to these types of data.  We address this challenge by including a binary component in our model specification (i.e., an indicator of a non-zero transmission probability between case pairs).  Finally, as is standard in most epidemiological analyses, our data do not include information on individuals who were exposed but not infected.  Therefore, the results should be interpreted conditional on both individuals being infected.  

While we use patristic distance as an outcome representative of symmetric dyadic genetic relatedness data in this work, the framework we have created can be readily adapted to SNP distances by modifying the likelihood in (1) to accommodate discrete count data.  In our R package \texttt{GenePair} (available at: \texttt{https://github.com/warrenjl/GenePair}), we have developed software for fitting the negative binomial and binary (i.e., genetically clustered or not) versions of the patristic distances model, along with the two versions described in Sections 3.1 and 3.2.  

Future work in this area could focus on alternative techniques for accounting for the correlation caused by analyzing dyadic outcomes while also incorporating spatial correlation.  Our current methods are based on the idea of shared spatially correlated random effect parameters between pairs of data including the same individual, which was shown to induce an intuitive correlation structure between observations in Section 3.3.  Additionally, these parameters were shown to be useful in identifying key individuals/areas that drive transmission in the study.  However, there are undoubtedly other approaches for achieving these same goals.  

A meta-regression approach could also be used in future work for genetic relatedness outcomes that are estimated in preceding analyses and include measures of uncertainty (something not applicable in our data).  The introduced random effects frameworks could still be used for defining the true but unobserved outcomes in this model, with a first stage that treats the observed outcome as an estimate of the true value.  A joint framework that combines the model which estimates the genetic relatedness outcomes with our models for characterizing variability and exploring associations in the outcomes could also represent an important extension.  However, the computational burden of this sort of approach would likely be great and could possibly require the addition of more efficient posterior sampling techniques. 

Overall, we recommend the use of the newly developed methods when the goal of a study is to estimate associations between spatially-referenced genetic relatedness and other variables of interest.  Even under the most simplistic, and likely unrealistic assumption of no correlation between dyadic responses, the new methodology performs well.  When correlation is present, competing approaches can yield potentially overly optimistic insights about the significance of the associations and/or poorly estimate the associations of interest and should not be used for decision making.

\bibliographystyle{chicago}
\bibliography{References}

\end{document}


\maketitle
\clearpage 

\doublespacing

\section{Induced correlation structure: Transmission probabilities}
In Figure S1, we present simulation-based correlation estimates for the transmission probabilities model over a range of spatial variance/correlation settings.  Using the same sample sizes, covariates, and spatial locations as observed in the Republic of Moldova dataset described in Section 2 of the main text, as well as parameter estimates obtained by fitting the model to those data in Section 5 (see Table S1), we simulate $1,000$ datasets from the full model detailed in Section 3.2 for each unique setting of $\phi$ and $\Omega$.  We vary $\phi$ from $0.0001$ to $10$ by increments of $0.10$ and multiply the estimate of $\Omega$ by a factor with the same range as $\phi$ to vary the amount of spatial correlation and variability, respectively, in the simulated responses.  

For a specific setting of these parameters, the $1,000$ simulated datasets are used to estimate the correlations of interest.  Specifically, we present estimates of the correlation between the $\pi_{ij}$ parameters, which are defined in (4) from the main text and control whether a transmission probability is larger than zero.  Correlations between these parameters are more robustly estimated than those between the raw outcomes or the $w_{ij}$ parameters given the large proportion of zeros in the real, and therefore simulated, data.  We present estimates of the correlation between $\pi_{ij}$ and $\pi_{ji}$ (i.e., two shared individuals, different roles), $\pi_{ij}$ and $\pi_{ik}$ (i.e., one shared individual, same role), $\pi_{ij}$ and $\pi_{ki}$ (i.e., one shared individual, different role), and $\pi_{ij}$ and $\pi_{kl}$ (i.e., no shared individual).
\clearpage 

\section{Simulation study details: Transmission probabilities}
\subsection{Data generation}
We simulate data from the distribution in (3) from the main text under three different scenarios.  In Setting 1, we specify that there is no unmeasured correlation by setting $\theta_{zi}^{\left(g\right)} = \theta_{zi}^{\left(r\right)} = \theta_{wi}^{\left(g\right)} = \theta_{wi}^{\left(r\right)} = \nu_{zi} = \nu_{wi} = 0$ for all $i$.  In Setting 2, we include random effect parameters but from (6) specify that $\eta_{z}^{\left(g\right)}\left(\textbf{s}^*_i\right) = \eta_{z}^{\left(r\right)}\left(\textbf{s}^*_i\right) = \eta_{w}^{\left(g\right)}\left(\textbf{s}^*_i\right) = \eta_{w}^{\left(r\right)}\left(\textbf{s}^*_i\right) = 0$ for all $i$, resulting in only non-spatial variability in the random effect parameters.  As a result, Setting 2 also ignores cross-covariance between parameters corresponding to the same individual.  In Setting 3, we also include random effect parameters but specify that $\phi = -\ln\left(0.05\right)/\max\left\{\left\|\textbf{s}_i^* - \textbf{s}_j^*\right\|; i < j\right\}$ and $\zeta_{zi}^{\left(g\right)} = \zeta_{zi}^{\left(r\right)} = \zeta_{wi}^{\left(g\right)} = \zeta_{wi}^{\left(r\right)} = 0$ for all $i$ in (6), resulting in multivariate spatial correlation that decreases to $0.05$ at the maximum distance observed in the data (i.e., the effective range) and non-zero cross-covariance.   

When simulating data from Setting 2, we define the total covariance matrix of the multivariate random effect process as $\text{diag}\left(\Omega\right) + \text{diag}\left(\sigma^2_{\zeta_z^{\left(g\right)}}, \sigma^2_{\zeta_z^{\left(r\right)}}, \sigma^2_{\zeta_w^{\left(g\right)}}, \sigma^2_{\zeta_w^{\left(r\right)}}\right)$ (i.e., no cross-covariance) and use the values in Table S1 to estimate this quantity.  For Setting 3, we consider the full covariance matrix as $\Omega + \text{diag}\left(\sigma^2_{\zeta_z^{\left(g\right)}}, \sigma^2_{\zeta_z^{\left(r\right)}}, \sigma^2_{\zeta_w^{\left(g\right)}}, \sigma^2_{\zeta_w^{\left(r\right)}}\right)$.  In Setting 2, all of this variability is assumed to be non-spatial while in Setting 3 it is attributed entirely to spatial correlation.  We generate a new vector of $\boldsymbol{\theta}$, $\boldsymbol{\nu}_{z}$, and $\boldsymbol{\nu}_{w}$ parameters for each dataset from the setting-specific model.  We also create a unique set of spatial locations, with no co-located individuals (i.e., $m = n$), for each dataset.  In total, we simulate 100 datasets from each setting.

\subsection{Competing models}
We apply three competing models to every dataset.  The first model (i.e., \textit{Fixed}) represents a simplified fixed effects regression form of (3-5) from the main text where $\theta_{zi}^{\left(g\right)} = \theta_{zi}^{\left(r\right)} = \theta_{wi}^{\left(g\right)} = \theta_{wi}^{\left(r\right)} = \nu_{zi} = \nu_{wi} = 0$ for all $i$ (i.e., no random effect parameters).  As \textit{Fixed} matches the data generated from Setting 1, we expect it to perform well overall in that setting.  However, in Settings 2 and 3, it will likely struggle in estimating the associations of interest as it ignores all correlation.  The second model (i.e., \textit{Non-spatial}) also represents a variant of the model (3-6) where $\eta_{z}^{\left(g\right)}\left(\textbf{s}^*_i\right) = \eta_{z}^{\left(r\right)}\left(\textbf{s}^*_i\right) = \eta_{w}^{\left(g\right)}\left(\textbf{s}^*_i\right) = \eta_{w}^{\left(r\right)}\left(\textbf{s}^*_i\right) = 0$ for all $i$ (matching data generation Setting 2).  Therefore, \textit{Non-spatial} accounts for correlation due to the nature of dyadic responses, but ignores the potential for spatial correlation and cross-covariance in the data.  It is currently unknown how inference for the regression parameters is impacted when dyadic correlation is accounted for but other correlation is ignored.  The final competing model (i.e., \textit{Spatial}) is our newly developed model in Section 3.2 which can address all sources of correlation simultaneously.    
\clearpage 

\section{Model fitting details}
We use Markov chain Monte Carlo sampling techniques (i.e., Gibbs and Metropolis-within-Gibbs algorithms) to fit the newly developed models within our R package \texttt{GenePair} \newline (available at: \texttt{https://github.com/warrenjl/GenePair}) \citep{metropolis1953equation, geman1984stochastic, gelfand1990sampling}.  In this section we present the full conditional distributions for all introduced model parameters needed to fit both of the models (i.e., patristic distances and transmission probabilities).

\subsection{Patristic distances}
The patristic distances model in Section 3.1 of the main text can be written more generally in the linear mixed model format such as $$\ln\left(\boldsymbol{P}\right) = X\boldsymbol{\delta} + Z\boldsymbol{\theta} + \boldsymbol{\epsilon}$$ where $\boldsymbol{\epsilon}|\sigma^2_{\epsilon} \sim \text{MVN}\left(\boldsymbol{0}_{n^*}, \sigma^2_{\epsilon}I_{n^*}\right)$; $n^* = \sum_{i=1}^{n-1} \sum_{j=i+1}^n 1$ is the total number of unique-pair patristic distances in the study; $n$ is the number of individuals in the study; $I_{n^*}$ is the $n^*$ by $n^*$ identity matrix; $\boldsymbol{P}^{\text{T}} = \left\{P_{12}, \hdots, P_{n-1,n}\right\}$ is the full vector of patristic distances (i.e., $P_{ij},\ i \neq j,\ i < j$); $X$ is an $n^*$ by $\left(p_x + p_d\right)$ matrix of covariates with $i^{th}$ row (corresponding to pair $(j,k)$ of data for example) equal to $\left\{\textbf{x}_{jk}^{\text{T}},  \left(\textbf{d}_j + \textbf{d}_k\right)^{\text{T}}\right\}$; $\boldsymbol{\delta}^{\text{T}} = \left(\boldsymbol{\beta}^{\text{T}}, \boldsymbol{\gamma}^{\text{T}}\right)$; $Z$ is an $n^*$ by $n$ matrix with $i^{th}$ row (corresponding to pair $(j,k)$ of data for example) equal to a vector of all zeros other than entries $j$ and $k$ which are equal to one; and $\boldsymbol{\theta}^{\text{T}} = \left(\theta_1, \hdots, \theta_n\right)$.

The prior distributions match those from Section 3.1.1 of the main text with \newline $\boldsymbol{\delta} \sim \text{MVN}\left(\boldsymbol{0}_{\left(p_x + p_d\right)}, 100^2 I_{\left(p_x + p_d\right)}\right)$ and $\boldsymbol{0}_p$ is a $p$-length vector of zeros; $\boldsymbol{\theta}|\boldsymbol{\eta}, \sigma^2_{\zeta} \sim \text{MVN}\left(V \boldsymbol{\eta}, \sigma^2_{\zeta} I_n\right)$ where $V$ is an $n$ by $m$ matrix with $V_{ij} = 1$ if individual $i$ resides at location $\textbf{s}^*_j$ and $V_{ij} = 0$ otherwise; $\boldsymbol{\eta}|\phi, \tau^2 \sim \text{MVN}\left\{\boldsymbol{0}_m, \tau^2 \Sigma\left(\phi\right)\right\}$; $\phi \sim \text{Gamma}\left(1.00, 1.00\right)$; and $\sigma^2_{\epsilon}, \sigma^2_{\zeta}, \tau^2 \stackrel{\text{iid}}{\sim} \text{Inverse Gamma}\left(0.01, 0.01\right)$.  For complete details on the definitions for each of these variables/parameters, please see Section 3.1.

Based on these specifications, the full conditional distributions are given as:
\begin{itemize}
    \item $\sigma^2_{\epsilon} | \Theta_{-\sigma^2_{\epsilon}}, \boldsymbol{P} \sim \text{Inverse Gamma}\left(\frac{n^*}{2} + 0.01, \frac{\left(\boldsymbol{P} - X\boldsymbol{\delta} - Z\boldsymbol{\theta}\right)^{\text{T}} \left(\boldsymbol{P} - X\boldsymbol{\delta} - Z\boldsymbol{\theta}\right)}{2} + 0.01\right),$ where $\Theta_{-\sigma^2_{\epsilon}}$ is the full vector of all parameters with $\sigma^2_{\epsilon}$ removed; 
    
    \item $\boldsymbol{\delta}|\Theta_{-\boldsymbol{\delta}}, \boldsymbol{P} \sim \text{MVN}\left(\Sigma_{\boldsymbol{\delta}}, \boldsymbol{\mu}_{\boldsymbol{\delta}}\right)$ with $\Sigma_{\boldsymbol{\delta}} = \left(\frac{X^{\text{T}}X}{\sigma^2_{\epsilon}} + \frac{I_{\left(p_x + p_d\right)}}{100^2}\right)^{-1}$ and $\boldsymbol{\mu}_{\boldsymbol{\delta}} = \frac{\Sigma_{\boldsymbol{\delta}} X^{\text{T}}\left(\boldsymbol{P} - Z\boldsymbol{\theta}\right)}{\sigma^2_{\epsilon}}$;
    
    \item $\boldsymbol{\theta}|\Theta_{-\boldsymbol{\theta}}, \boldsymbol{P} \sim \text{MVN}\left(\Sigma_{\boldsymbol{\theta}}, \boldsymbol{\mu}_{\boldsymbol{\theta}}\right)$ with $\Sigma_{\boldsymbol{\theta}} = \left(\frac{Z^{\text{T}}Z}{\sigma^2_{\epsilon}} + \frac{I_n}{\sigma^2_{\zeta}}\right)^{-1}$ and $\boldsymbol{\mu}_{\boldsymbol{\theta}} = \Sigma_{\boldsymbol{\theta}} \left\{ \frac{Z^{\text{T}}\left(\boldsymbol{P} - X\boldsymbol{\delta}\right)}{\sigma^2_{\epsilon}} + \frac{V\boldsymbol{\eta}}{\sigma^2_{\zeta}}\right\}$; we implement a sum-to-zero constraint \textit{on the fly} for these parameters \citep{besag1995bayesian, berrocal2012space, warren2021spatial}; 
    
    \item $\sigma^2_{\zeta} | \Theta_{-\sigma^2_{\zeta}}, \boldsymbol{P} \sim \text{Inverse Gamma}\left(\frac{n}{2} + 0.01, \frac{\left(\boldsymbol{\theta} - V\boldsymbol{\eta}\right)^{\text{T}} \left(\boldsymbol{\theta} - V\boldsymbol{\eta}\right)}{2} + 0.01\right)$;
    
    \item $\boldsymbol{\eta}|\Theta_{-\boldsymbol{\eta}}, \boldsymbol{P} \sim \text{MVN}\left(\Sigma_{\boldsymbol{\eta}}, \boldsymbol{\mu}_{\boldsymbol{\eta}}\right)$ with $\Sigma_{\boldsymbol{\eta}} = \left(\frac{V^{\text{T}}V}{\sigma^2_{\zeta}} + \frac{\Sigma\left(\phi\right)^{-1}}{\tau^2}\right)^{-1}$ and $\boldsymbol{\mu}_{\boldsymbol{\eta}} = \frac{\Sigma_{\boldsymbol{\eta}} V^{\text{T}}\boldsymbol{\theta}}{\sigma^2_{\zeta}}$;
    
    \item $\tau^2 | \Theta_{-\tau^2}, \boldsymbol{P} \sim \text{Inverse Gamma}\left(\frac{m}{2} + 0.01, \frac{\boldsymbol{\eta}^{\text{T}} \Sigma\left(\phi\right)^{-1} \boldsymbol{\eta}}{2} + 0.01\right)$; and
    
    \item $f\left\{\ln\left(\phi\right)| \Theta_{-\phi}, \boldsymbol{P}\right\} \propto \left|\Sigma\left(\phi\right)^{-1}\right|^{1/2} \exp\left\{-\frac{1}{2\tau^2} \boldsymbol{\eta}^{\text{T}} \Sigma\left(\phi\right)^{-1} \boldsymbol{\eta}\right\}f\left\{\ln\left(\phi\right)\right\}$; during model fitting, we work with $\ln\left(\phi\right) \in \mathbb{R}$ instead of $\phi$ so that a symmetric proposal density can be used in the Metropolis algorithm. 
    
\end{itemize}

\subsection{Transmission probabilities}
From the transmission probabilities model introduced in Section 3.2 of the main text, we work with the complete vector of observed transmission probabilities, $\boldsymbol{T}^{\text{T}} = \left(T_{12}, \hdots, T_{1n}, T_{21}, T_{23}, \hdots, T_{n,n-1}\right)$, whose length is $n^* = \sum_{i=1}^n \sum_{j=1, i \neq j}^n 1$ (i.e., $T_{ij}$ and $T_{ji}$ included for $i \neq j$).  We also introduce a vector of latent probabilities controlling whether the transmission probabilities are $> 0$ (i.e., $\boldsymbol{\pi}$) and a vector of latent continuous variables representing the logit-scaled transmission probabilities (i.e., $\boldsymbol{w}$), both of which are the same length and ordered in the same way as $\boldsymbol{T}$.  We note that not every $T_{ij} > 0$, meaning that not every $w_{ij}$ is directly observed.  When $T_{ij} = 0$, the corresponding $w_{ij}$ is treated as a missing variable within this framework.  

We then use the regression frameworks defined in Section 3.2 of the main text to define these latent parameter vectors such that \begin{align*} &\text{logit} \left(\boldsymbol{\pi}\right) = X\boldsymbol{\delta}_z + Z^{(g)}\boldsymbol{\theta}_z^{(g)} + Z^{(r)}\boldsymbol{\theta}_z^{(r)} + \boldsymbol{\nu}_z^* \text{ and}\\ &\boldsymbol{w} = X\boldsymbol{\delta}_w + Z^{(g)}\boldsymbol{\theta}_w^{(g)} + Z^{(r)}\boldsymbol{\theta}_w^{(r)} + \boldsymbol{\nu}_w^* + \boldsymbol{\epsilon} \end{align*} where $\boldsymbol{\epsilon}|\sigma^2_{\epsilon} \sim \text{MVN}\left(\boldsymbol{0}_{n^*}, \sigma^2_{\epsilon}I_{n^*}\right)$; $X$ is an $n^*$ by $(p_x + 2p_d)$ matrix of covariates with $i^{th}$ row (corresponding to pair $(j,k)$ of data for example) equal to $\left\{\textbf{x}_{jk}^{\text{T}},\  \textbf{d}_k^{\text{T}},\ \textbf{d}_j^{\text{T}}\right\}$; $\boldsymbol{\delta}_z^{\text{T}} = \left(\boldsymbol{\beta}_z^{\text{T}}, \ \boldsymbol{\gamma}_z^{(g)\text{T}},\ \boldsymbol{\gamma}_z^{(r)\text{T}}\right)$ with $\boldsymbol{\delta}_w$ defined similarly; $Z^{(g)}$ is an $n^*$ by $n$ matrix with $i^{th}$ row (corresponding to pair $(j,k)$ of data for example) equal to a vector of all zeros other than entry $k$ which is equal to one; $Z^{(r)}$ is defined similarly to $Z^{(g)}$ where entry $j$ is equal to one instead of entry $k$; $\boldsymbol{\theta}_z^{(g)\text{T}} = \left(\theta_{z1}^{(g)}, \hdots, \theta_{zn}^{(g)}\right)$ with $\boldsymbol{\theta}_z^{(r)}$, $\boldsymbol{\theta}_w^{(g)}$, and $\boldsymbol{\theta}_w^{(r)}$ defined similarly; and $\boldsymbol{\nu}^{*\text{T}}_z = \left(\nu_{z1}\nu_{z2}, \hdots, \nu_{z1}\nu_{zn}, \nu_{z2}\nu_{z1}, \nu_{z2}\nu_{z3}, \hdots, \nu_{zn}\nu_{z,n-1}\right)$ is the vector of interaction terms with $\boldsymbol{\nu}^*_w$ defined similarly. 

The prior distributions match those from Section 3.2.1 of the main text with \newline $\boldsymbol{\delta}_z, \boldsymbol{\delta}_w \stackrel{\text{iid}}{\sim} \text{MVN}\left(\boldsymbol{0}_{\left(p_x + 2p_d\right)}, 100^2 I_{\left(p_x + 2p_d\right)}\right)$; $\boldsymbol{\theta}_z^{(g)} | \boldsymbol{\eta}_z^{(g)}, \sigma^2_{\zeta_z^{(g)}} \sim \text{MVN}\left(V \boldsymbol{\eta}_z^{(g)}, \sigma^2_{\zeta_z^{(g)}} I_n\right)$ where $\boldsymbol{\theta}_z^{(r)}$, $\boldsymbol{\theta}_w^{(g)}$, and $\boldsymbol{\theta}_w^{(r)}$ are defined similarly; $\sigma^2_{\epsilon}, \sigma^2_{\zeta_z^{(g)}}, \sigma^2_{\zeta_z^{(r)}}, \sigma^2_{\zeta_w^{(g)}}, \sigma^2_{\zeta_w^{(r)}}, \sigma^2_{\nu_z}, \sigma^2_{\nu_w} \stackrel{\text{iid}}{\sim} \text{Inverse Gamma}\left(0.01, 0.01\right)$; \newline $\left(\boldsymbol{\eta}_z^{(g)\text{T}}, \boldsymbol{\eta}_z^{(r)\text{T}}, \boldsymbol{\eta}_w^{(g)\text{T}}, \boldsymbol{\eta}_w^{(r)\text{T}} \right)^{\text{T}} | \phi, \Omega \sim \text{MVN}\left\{\boldsymbol{0}_{4m}, \Omega \otimes \Sigma\left(\phi\right)\right\}$; $\phi \sim \text{Gamma}\left(1.00, 1.00\right)$; and $\Sigma^{-1} \sim \text{Wishart}\left(5, I_4\right)$.  For complete details on the definitions for each of these variables/parameters, please see Section 3.2.

Because we are working in a logistic regression framework, we use the P\'olya-Gamma latent variable approach for posterior sampling developed by \cite{polson2013bayesian}.  Specifically, for each $T_{ij}$ outcome, we introduce a corresponding P\'olya-Gamma distributed latent variable such that $$\omega^*_{ij}| \boldsymbol{\delta}_z, \theta_{zj}^{(g)}, \theta_{zi}^{(r)} \stackrel{\text{ind}}{\sim} \text{P\'olya-Gamma}\left(1, \textbf{x}_{ij}^{\text{T}}\boldsymbol{\beta}_z + \textbf{d}_j^{\text{T}}\boldsymbol{\gamma}_z^{(g)} + \textbf{d}_i^{\text{T}}\boldsymbol{\gamma}_z^{(r)}  + \theta_{zj}^{(g)} + \theta_{zi}^{(r)} + \nu_{zi}\nu_{zj}\right).$$  These latent variables allow for closed form full conditional distributions for several of the parameter sets and we sample from their distribution using the \texttt{pgdraw} package in R \citep{pgdraw}. 

Based on these specifications, the full conditional distributions are given as:
    
    \begin{itemize}
    
    \item $\omega^*_{ij} | \boldsymbol{\Theta}_{-\omega^*_{ij}}, \boldsymbol{T} \stackrel{\text{ind}}{\sim} \text{P\'olya-Gamma}\left(1, \textbf{x}_{ij}^{\text{T}}\boldsymbol{\beta}_z + \textbf{d}_j^{\text{T}}\boldsymbol{\gamma}_z^{(g)} + \textbf{d}_i^{\text{T}}\boldsymbol{\gamma}_z^{(r)}  + \theta_{zj}^{(g)} + \theta_{zi}^{(r)} + \nu_{zi}\nu_{zj}\right)$;
    
     \item $\boldsymbol{\delta}_z|\Theta_{-\boldsymbol{\delta}_z}, \boldsymbol{T} \sim \text{MVN}\left(\Sigma_{\boldsymbol{\delta}_z}, \boldsymbol{\mu}_{\boldsymbol{\delta}_z}\right)$ with $\Sigma_{\boldsymbol{\delta}_z} = \left(X^{\text{T}} \Omega^* X + \frac{I_{\left(p_x + 2p_d\right)}}{100^2}\right)^{-1}$ and \\ $\boldsymbol{\mu}_{\boldsymbol{\delta}_z} = \Sigma_{\boldsymbol{\delta}_z} X^{\text{T}} \Omega^* \left(\boldsymbol{\lambda} - Z^{(g)}\boldsymbol{\theta}_z^{(g)} - Z^{(r)}\boldsymbol{\theta}_z^{(r)} - \boldsymbol{\nu}^*_z\right)$, where $\Omega^*$ is an $n^*$ by $n^*$ diagonal matrix with the diagonal vector equal to $\boldsymbol{\omega}^{*\text{T}} = \left(\omega^*_{12}, \hdots, \omega^*_{n, n-1}\right)$ and \newline $\boldsymbol{\lambda}^{\text{T}} = \left(\frac{1\left(T_{12} > 0\right) - 0.50}{\omega^*_{12}}, \hdots, \frac{1\left(T_{n,n-1} > 0\right) - 0.50}{\omega^*_{n,n-1}}\right)$;
     
    \item $f\left(\nu_{zi} | \Theta_{-\nu_{zi}}, \boldsymbol{T}\right) \propto f\left(\boldsymbol{T}|\boldsymbol{\Theta}\right)f\left(\nu_{zi} | \sigma^2_{\nu_z}\right)$, where $f\left(\boldsymbol{T}|\boldsymbol{\Theta}\right)$ is the joint distribution of the data and we use a Metropolis algorithm to sample from this distribution;
    
    \item $\sigma^2_{\nu_z} | \Theta_{-\sigma^2_{\nu_z}}, \boldsymbol{T} \sim \text{Inverse Gamma}\left(\frac{n}{2} + 0.01, \frac{\boldsymbol{\nu}_z^{\text{T}} \boldsymbol{\nu}_z}{2} + 0.01\right)$, where $\boldsymbol{\nu}_z^{\text{T}} = \left(\nu_{z1}, \hdots, \nu_{zn}\right)$;
     
     \item $\boldsymbol{\theta}_z^{(g)}|\Theta_{-\boldsymbol{\theta}_z^{(g)}}, \boldsymbol{T} \sim \text{MVN}\left(\Sigma_{\boldsymbol{\theta}_z^{(g)}}, \boldsymbol{\mu}_{\boldsymbol{\theta}_z^{(g)}}\right)$ with $\Sigma_{\boldsymbol{\theta}_z^{(g)}} = \left(Z^{(g)\text{T}} \Omega^* Z^{(g)} + \frac{I_n}{\sigma^2_{\zeta_z^{(g)}}}\right)^{-1}$ and $\boldsymbol{\mu}_{\boldsymbol{\theta}_z^{(g)}} = \Sigma_{\boldsymbol{\theta}_z^{(g)}} \left\{Z^{(g)\text{T}} \Omega^* \left(\boldsymbol{\lambda} - X\boldsymbol{\delta}_z - Z^{(r)}\boldsymbol{\theta}_z^{(r)} - \boldsymbol{\nu}^*_w\right) + \frac{V \boldsymbol{\eta}_z^{(g)}}{\sigma^2_{\zeta_z^{(g)}}}\right\}$; we implement a sum-to-zero constraint \textit{on the fly} for these parameters \citep{besag1995bayesian, berrocal2012space, warren2021spatial}; 
     
     \item $f\left(\boldsymbol{\theta}_z^{(r)}|\Theta_{-\boldsymbol{\theta}_z^{(r)}}, \boldsymbol{T}\right)$ has similar form to $f\left(\boldsymbol{\theta}_z^{(g)}|\Theta_{-\boldsymbol{\theta}_z^{(g)}}, \boldsymbol{T}\right)$ with $(g)$ replaced by $(r)$;
     
     \item $\sigma^2_{\zeta_z^{(g)}} | \Theta_{-\sigma^2_{\zeta_z^{(g)}}}, \boldsymbol{T} \sim \text{Inverse Gamma}\left(\frac{n}{2} + 0.01, \frac{\left(\boldsymbol{\theta}_z^{(g)} - V \boldsymbol{\eta}_z^{(g)}\right)^{\text{T}} \left(\boldsymbol{\theta}_z^{(g)} - V \boldsymbol{\eta}_z^{(g)}\right)}{2} + 0.01\right)$;
     
     \item $f\left(\sigma^2_{\zeta_z^{(r)}} | \Theta_{-\sigma^2_{\zeta_z^{(r)}}}, \boldsymbol{T}\right)$, $f\left(\sigma^2_{\zeta_w^{(g)}} | \Theta_{-\sigma^2_{\zeta_w^{(g)}}}, \boldsymbol{T}\right)$, and $f\left(\sigma^2_{\zeta_w^{(r)}} | \Theta_{-\sigma^2_{\zeta_w^{(r)}}}, \boldsymbol{T}\right)$ each have the same form as $f\left(\sigma^2_{\zeta_z^{(g)}} | \Theta_{-\sigma^2_{\zeta_z^{(g)}}}, \boldsymbol{T}\right)$ with $(g)$ replaced by $(r)$ and $w$ replaced by $z$ when appropriate;
     
     \item  $\boldsymbol{\eta}_z^{(g)}|\Theta_{-\boldsymbol{\eta}_z^{(g)}}, \boldsymbol{T} \sim \text{MVN}\left(\Sigma_{\boldsymbol{\eta}_z^{(g)}}, \boldsymbol{\mu}_{\boldsymbol{\eta}_z^{(g)}}\right)$ with $\Sigma_{\boldsymbol{\eta}_z^{(g)}} = \left\{\frac{V^{\text{T}}V}{\sigma^2_{\zeta_z^{(g)}}} + \left(\Sigma_{11} - \Sigma_{12}\Sigma_{22}^{-1}\Sigma_{21}\right)^{-1}\right\}^{-1}$ and $\boldsymbol{\mu}_{\boldsymbol{\eta}_z^{(g)}} = \Sigma_{\boldsymbol{\eta}_z^{(g)}} \left\{\frac{V^{\text{T}}\boldsymbol{\theta}_z^{(g)}}{\sigma^2_{\zeta_z^{(g)}}} + \left(\Sigma_{11} - \Sigma_{12}\Sigma_{22}^{-1}\Sigma_{12}\right)^{-1} \Sigma_{12}\Sigma_{22}^{-1} \begin{bmatrix}
           \boldsymbol{\eta}_z^{(r)} \\
           \boldsymbol{\eta}_w^{(g)} \\
           \boldsymbol{\eta}_w^{(r)}
         \end{bmatrix}\right\}$, where $\Sigma_{11} = \Omega_{11} \Sigma\left(\phi\right)$, $\Sigma_{12} = \left\{\Omega_{12}\Sigma\left(\phi\right), \Omega_{13}\Sigma\left(\phi\right), \Omega_{14}\Sigma\left(\phi\right)\right\}$, $\Sigma_{21} = \Sigma_{12}^{\text{T}}$, and $\Sigma_{22} = \Omega_{2:4,2:4} \otimes \Sigma\left(\phi\right)$;
         
    \item $f\left(\boldsymbol{\eta}_z^{(r)}|\Theta_{-\boldsymbol{\eta}_z^{(r)}}, \boldsymbol{T}\right)$, $f\left(\boldsymbol{\eta}_w^{(g)}|\Theta_{-\boldsymbol{\eta}_w^{(g)}}, \boldsymbol{T}\right)$, and $f\left(\boldsymbol{\eta}_w^{(r)}|\Theta_{-\boldsymbol{\eta}_w^{(r)}}, \boldsymbol{T}\right)$ each have a similar form to $f\left(\boldsymbol{\eta}_z^{(g)}|\Theta_{-\boldsymbol{\eta}_z^{(g)}}, \boldsymbol{T}\right)$ with the obvious $(g)$, $(r)$, $z$, $w$ changes and updates of the subscripts on the entries of $\Omega$ and ordering of the vectors;
    
    \item $w_{ij} | \Theta_{-w_{ij}}, \boldsymbol{T}_{\left\{T_{ij} > 0\right\}} \equiv \text{logit}\left(T_{ij}\right)$;
    
    \item $w_{ij} | \Theta_{-w_{ij}}, \boldsymbol{T}_{\left\{T_{ij} = 0\right\}} \stackrel{\text{ind}}{\sim} \text{N}\left(\textbf{x}_{ij}^{\text{T}}\boldsymbol{\beta}_w + \textbf{d}_j^{\text{T}}\boldsymbol{\gamma}_w^{(g)} + \textbf{d}_i^{\text{T}}\boldsymbol{\gamma}_w^{(r)}  + \theta_{wj}^{(g)} + \theta_{wi}^{(r)} + \nu_{wi}\nu_{wj}, \sigma^2_{\epsilon}\right)$;
     
    \item $\boldsymbol{\delta}_w|\Theta_{-\boldsymbol{\delta}_w}, \boldsymbol{T} \sim \text{MVN}\left(\Sigma_{\boldsymbol{\delta}_w}, \boldsymbol{\mu}_{\boldsymbol{\delta}_w}\right)$ with $\Sigma_{\boldsymbol{\delta}_w} = \left(\frac{X^{\text{T}}X}{\sigma^2_{\epsilon}} + \frac{I_{\left(p_x + 2p_d\right)}}{100^2}\right)^{-1}$ and \\ $\boldsymbol{\mu}_{\boldsymbol{\delta}_w} = \frac{\Sigma_{\boldsymbol{\delta}_w}  X^{\text{T}} \left(\boldsymbol{w} - Z^{(g)}\boldsymbol{\theta}_w^{(g)} - Z^{(r)}\boldsymbol{\theta}_w^{(r)} - \boldsymbol{\nu}^*_w\right)}{\sigma^2_{\epsilon}}$;
    
    \item $\sigma^2_{\epsilon} | \Theta_{-\sigma^2_{\epsilon}}, \boldsymbol{T} \sim \newline \text{Inverse Gamma}\left(\frac{n^*}{2} + 0.01, \frac{\left(\boldsymbol{w} - X\boldsymbol{\delta}_w - Z^{(g)}\boldsymbol{\theta}_w^{(g)} - Z^{(r)}\boldsymbol{\theta}_w^{(r)} - \boldsymbol{\nu}^*_w\right)^{\text{T}} \left(\boldsymbol{w} - X\boldsymbol{\delta}_w - Z^{(g)}\boldsymbol{\theta}_w^{(g)} - Z^{(r)}\boldsymbol{\theta}_w^{(r)} - \boldsymbol{\nu}^*_w\right)}{2} + 0.01\right)$;
    
    \item $f\left(\nu_{wi} | \Theta_{-\nu_{wi}}, \boldsymbol{T}\right) \propto f\left(\boldsymbol{T}|\boldsymbol{\Theta}\right)f\left(\nu_{wi} | \sigma^2_{\nu_w}\right)$, where $f\left(\boldsymbol{T}|\boldsymbol{\Theta}\right)$ is the joint distribution of the data and we use a Metropolis algorithm to sample from this distribution;
        
    \item $\sigma^2_{\nu_w} | \Theta_{-\sigma^2_{\nu_w}}, \boldsymbol{T} \sim \text{Inverse Gamma}\left(\frac{n}{2} + 0.01, \frac{\boldsymbol{\nu}_w^{\text{T}} \boldsymbol{\nu}_w}{2} + 0.01\right)$, where $\boldsymbol{\nu}_w^{\text{T}} = \left(\nu_{w1}, \hdots, \nu_{wn}\right)$;
        
    \item $\boldsymbol{\theta}_w^{(g)}|\Theta_{-\boldsymbol{\theta}_w^{(g)}}, \boldsymbol{T} \sim \text{MVN}\left(\Sigma_{\boldsymbol{\theta}_w^{(g)}}, \boldsymbol{\mu}_{\boldsymbol{\theta}_w^{(g)}}\right)$ with $\Sigma_{\boldsymbol{\theta}_w^{(g)}} = \left(\frac{Z^{(g)\text{T}}Z^{(g)}}{\sigma^2_{\epsilon}} + \frac{I_n}{\sigma^2_{\zeta_w^{(g)}}}\right)^{-1}$ and \newline $\boldsymbol{\mu}_{\boldsymbol{\theta}_w^{(g)}} = \Sigma_{\boldsymbol{\theta}_w^{(g)}} \left\{ \frac{Z^{(g)\text{T}}\left(\boldsymbol{w} - X\boldsymbol{\delta}_w - Z^{(r)}\boldsymbol{\theta}_w^{(r)} - \boldsymbol{\nu}^*_w\right)}{\sigma^2_{\epsilon}} + \frac{V\boldsymbol{\eta}_w^{(g)}}{\sigma^2_{\zeta_w^{(g)}}}\right\}$; we implement a sum-to-zero constraint \textit{on the fly} for these parameters; 
    
    \item $f\left(\boldsymbol{\theta}_w^{(r)}|\Theta_{-\boldsymbol{\theta}_w^{(r)}}, \boldsymbol{T}\right)$ has the same as $f\left(\boldsymbol{\theta}_w^{(g)}|\Theta_{-\boldsymbol{\theta}_w^{(g)}}, \boldsymbol{T}\right)$ with $(g)$ replaced by $(r)$;
    
    \item $\Omega^{-1}|\Theta_{-\Omega^{-1}}, \boldsymbol{T} \sim \text{Wishart}\left\{\left(\sum_{j=1}^m \sum_{i=1}^m \boldsymbol{\eta}\left(\textbf{s}^*_j\right) \boldsymbol{\eta}\left(\textbf{s}^*_i\right)^{\text{T}} \Sigma\left(\phi\right)^{-1}_{ij} + I_4\right)^{-1}, m + 5\right\}$; and
    
    \item $f\left\{\ln\left(\phi\right)| \Theta_{-\phi}, \boldsymbol{T}\right\} \propto \left|\Sigma\left(\phi\right)^{-1}\right|^{2} \exp\left[-\frac{1}{2} \boldsymbol{\eta}^{\text{T}} \left\{\Sigma\left(\phi\right)^{-1} \otimes \Omega^{-1}\right\} \boldsymbol{\eta}\right]f\left\{\ln\left(\phi\right)\right\}$; as in Section S3.1, we work with $\ln\left(\phi\right) \in \mathbb{R}$ instead of $\phi$ so that a symmetric proposal density can be used in the Metropolis algorithm.
    
    \end{itemize}

\clearpage 

\section{Random effect parameter analyses}
\subsection{Summarizing the estimates}
We make inference on the random effect parameters and hyperparameters from both the patristic distance and transmission probability models and summarize them with respect to the raw genetic distance outcomes.  Results are shown in Table S2.  For the patristic distance results, we observe that 57 of the 99 $\theta_i$ parameters have CIs that exclude zero; 29 of which are positive and 28 that are negative.  Recall that individuals with positive $\theta_i$ are more likely to be in pairs with larger patristic distances (i.e., less likely transmission) while those with negative $\theta_i$ are more likely to be in pairs with small patristic distances (i.e., transmission pair).  To test this, we calculate the average patristic distance for all outcome pairs that involve the individuals with significantly positive $\theta_i$, and repeat the calculation for those with significantly negative $\theta_i$.  For reference, the average patristic distance overall (multiplied by 10,000 for presentation purposes) is 3.12 while for the positive and negative random effect parameter pairs is 3.96 and 2.47, respectively.  This shows the intuitive connection between the raw data and estimated values of these parameters, and the ability of the framework to identify key drivers of transmission in the population.  Around 23\% (5\%, 68\%) of the total variability in the $\theta_i$ parameters is spatially structured based on analyzing the posterior distribution of $\tau^2/\left(\tau^2 + \sigma^2_{\zeta}\right)$.  

For the transmission probability outcome results, as expected individuals identified with significant, positive random effect parameters (both giver and receiver type) are more likely to have a larger, non-zero transmission probability.  Also, the proportion of total variability in the random effects due to the multivariate spatial process is much higher with point estimates ranging from 74\% to 98\%.  

\subsection{Spatial prediction}
In Figures S2 and S3, we display posterior means and standard deviations, respectively, for the random effect parameters predicted across the Republic of Moldova.  To produce these maps, we create an equally spaced grid over the domain containing 500 new locations (i.e., where data were not previously observed) and then collect samples from the posterior predictive distribution of the random effect parameters at each location separately (e.g., samples from $f\left\{\theta_{0}|\ln\left(P_{12}\right),\hdots,\ln\left(P_{n-1,n}\right)\right\}$ for the patristic distances model) using techniques from \cite{banerjee2014hierarchical}.  These samples are then summarized using posterior means and standard deviations and plotted across the domain using inverse distance weighting to fill in the remaining gaps (for visualization purposes only).  Areas in red from the posterior mean map suggest that patristic distances/transmission probabilities involving individuals residing in these locations tend to be smaller/larger on average than areas in blue, possibly indicating geographic areas of increased residual transmission activity.  For patristic distances however, the variability in these values overall is not large in this setting, suggesting that the risk does not differ substantially across the map.

\clearpage 
\section{Additional figures and tables}
\begin{figure}[h!]
\begin{center}
\includegraphics[trim={0.0cm 0.0cm 3.8cm 0.0cm}, clip, scale = 0.29]{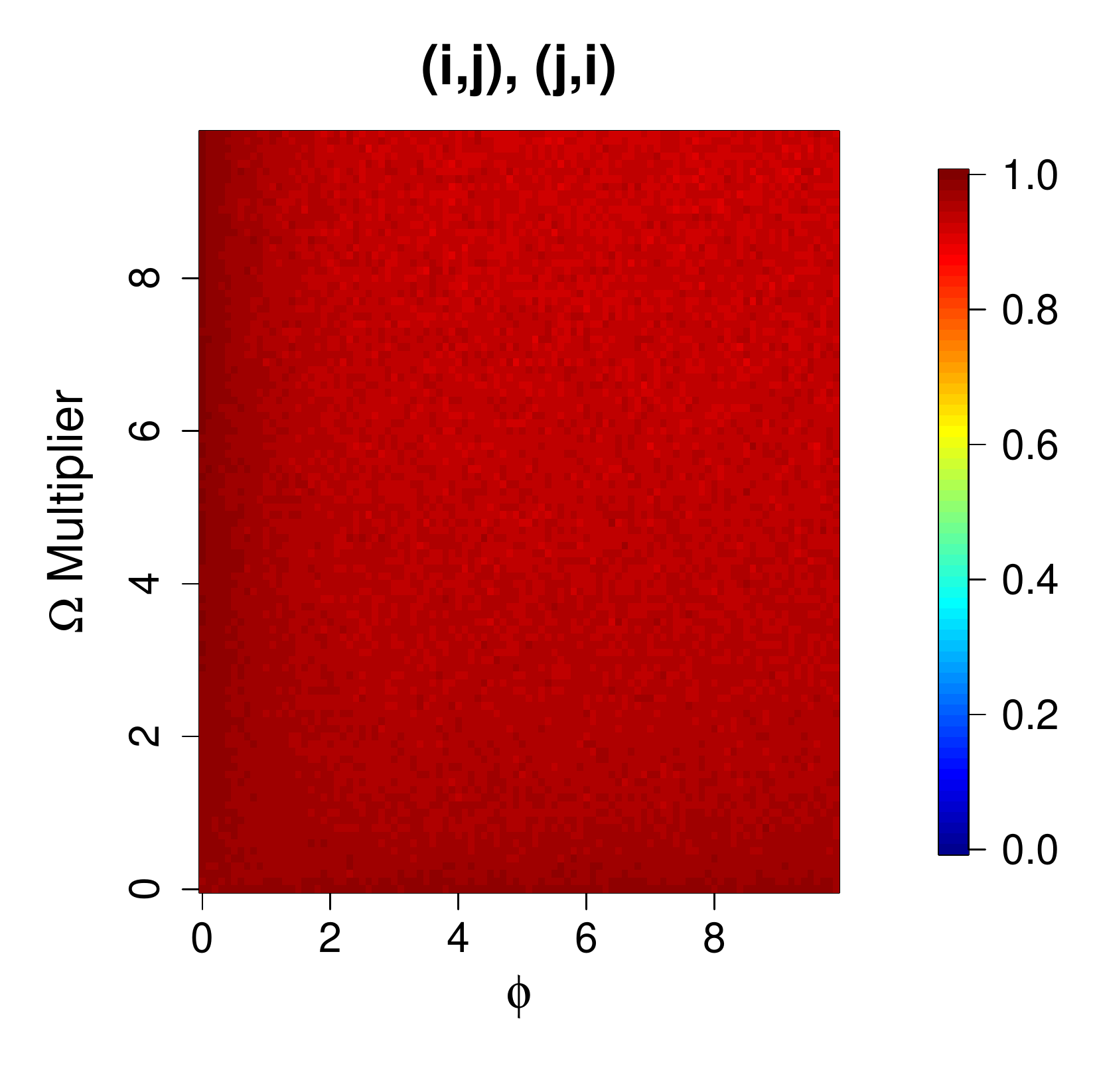}
\includegraphics[trim={0.0cm 0.0cm 3.8cm 0.0cm}, clip, scale = 0.29]{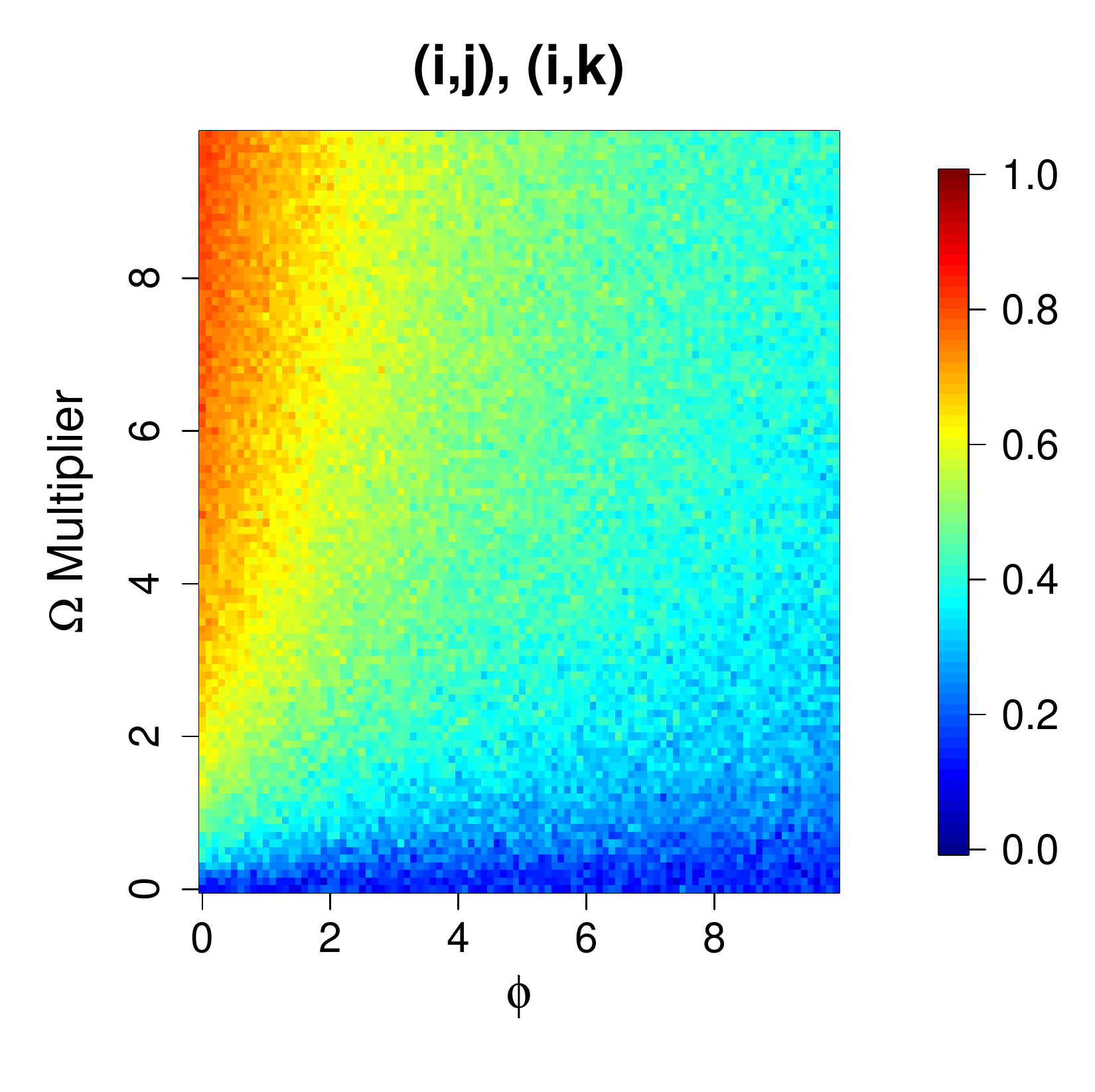}
\includegraphics[trim={0.0cm 0.0cm 3.8cm 0.0cm}, clip, scale = 0.29]{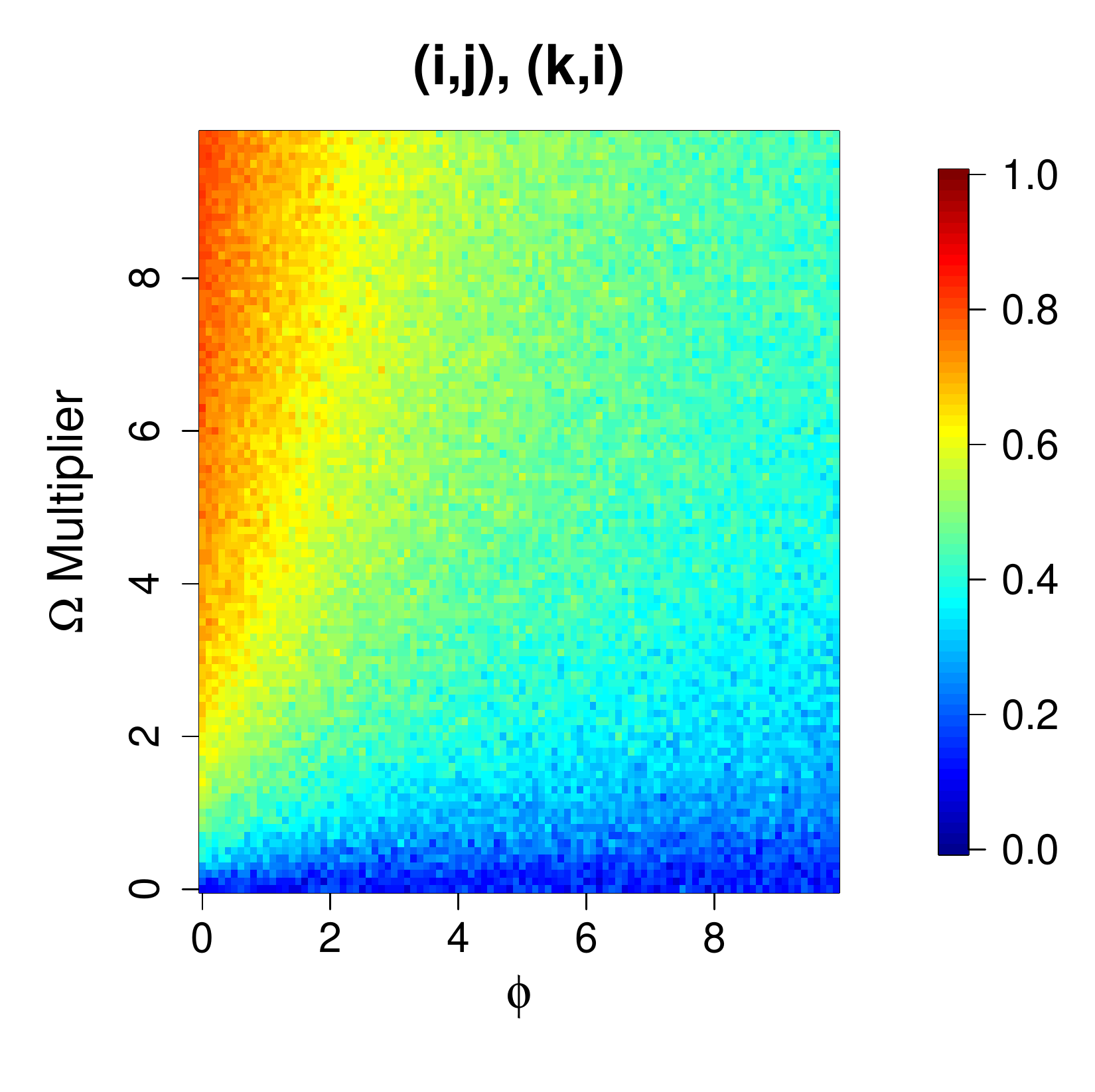}
\includegraphics[scale = 0.29]{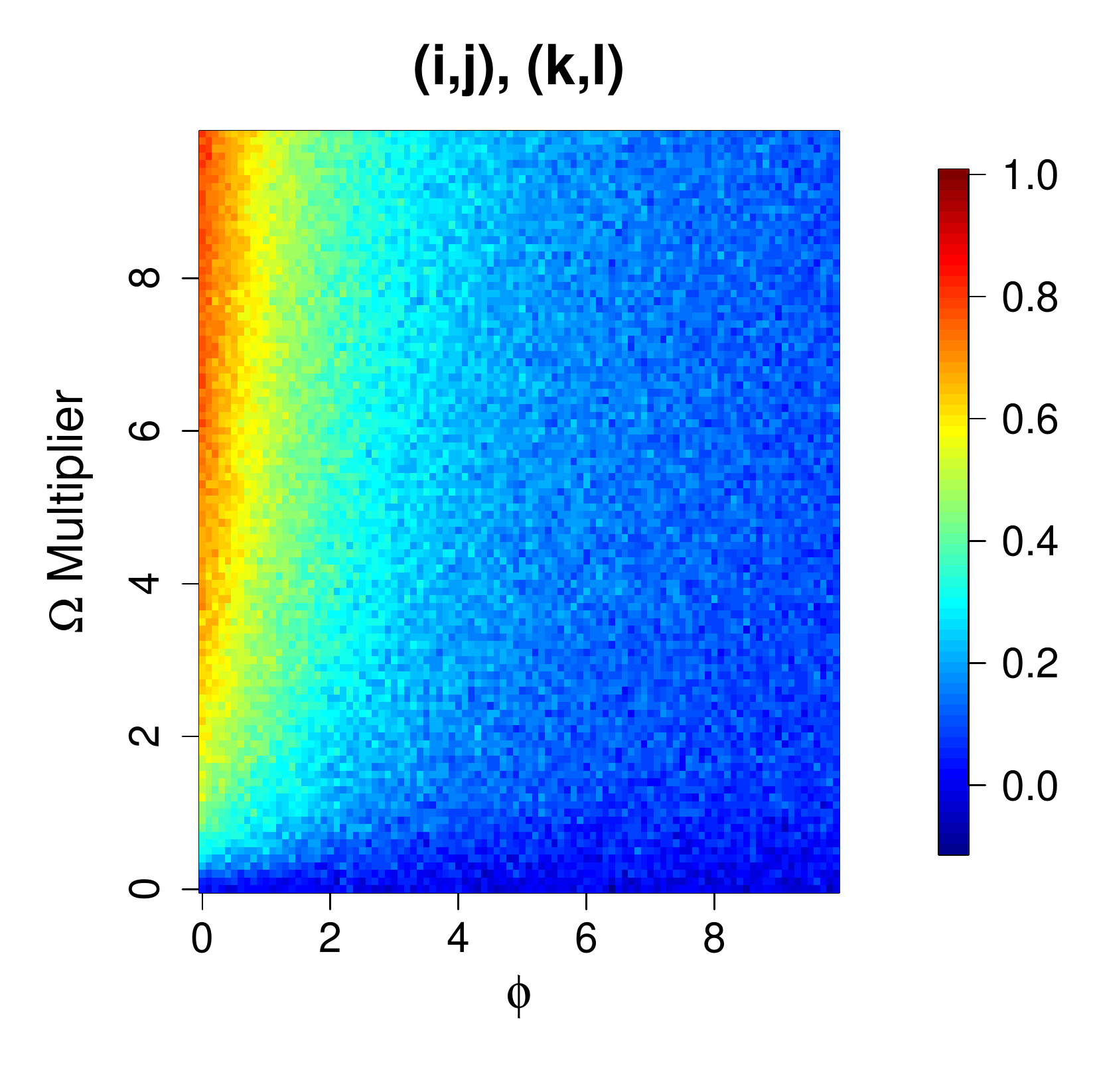}
\caption{Simulation-based correlation estimates between the $\pi_{ij}$ parameters from the transmission probabilities model for different pairs of individuals.}
\end{center}
\end{figure}
\clearpage 

\begin{figure}[ht]
\centering
\includegraphics[trim={0.0cm 0.0cm 0.0cm 1.0cm}, clip, scale = 0.30]{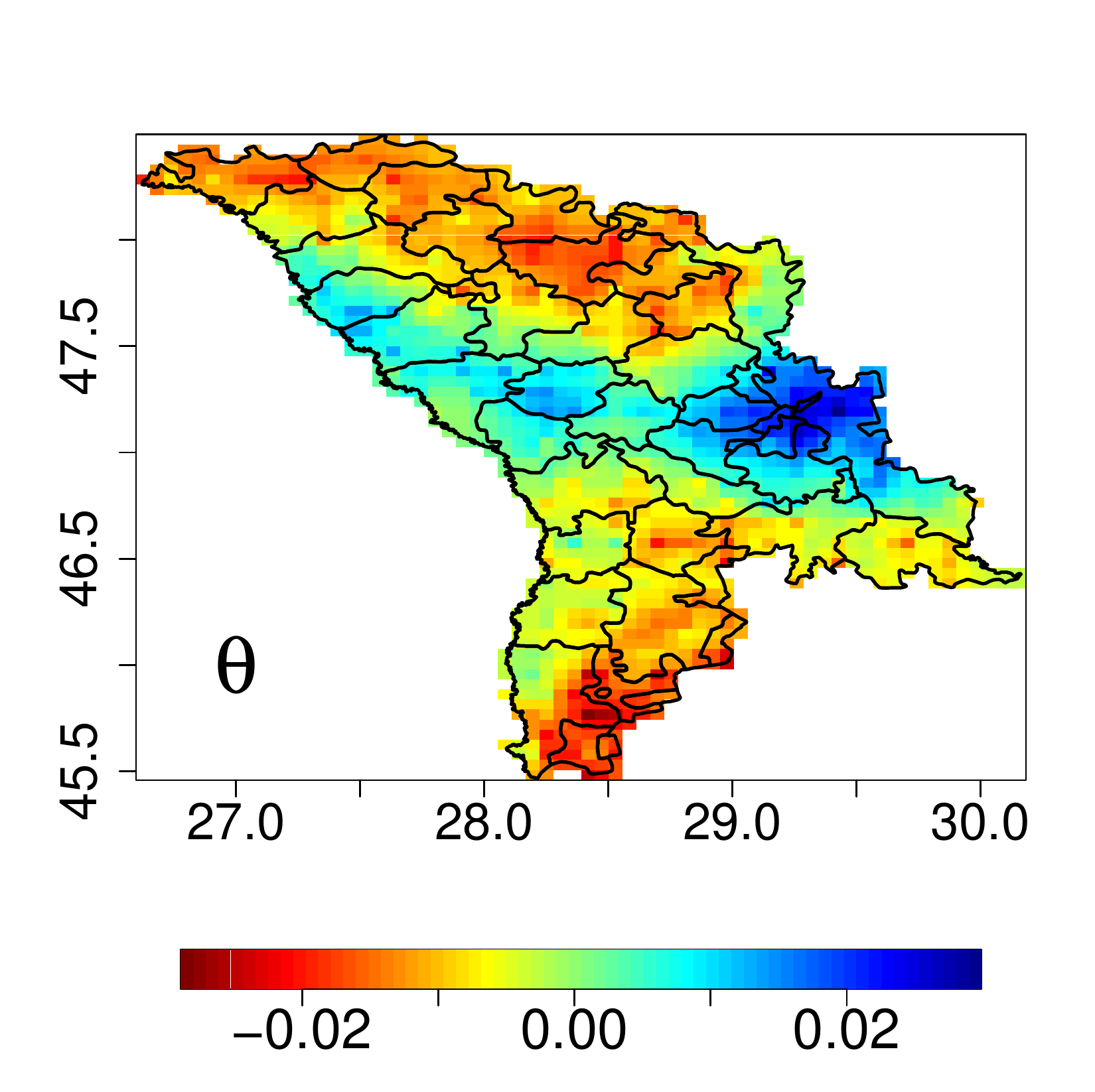}\\
\includegraphics[trim={0.0cm 0.0cm 0.0cm 1.0cm}, clip, scale = 0.30]{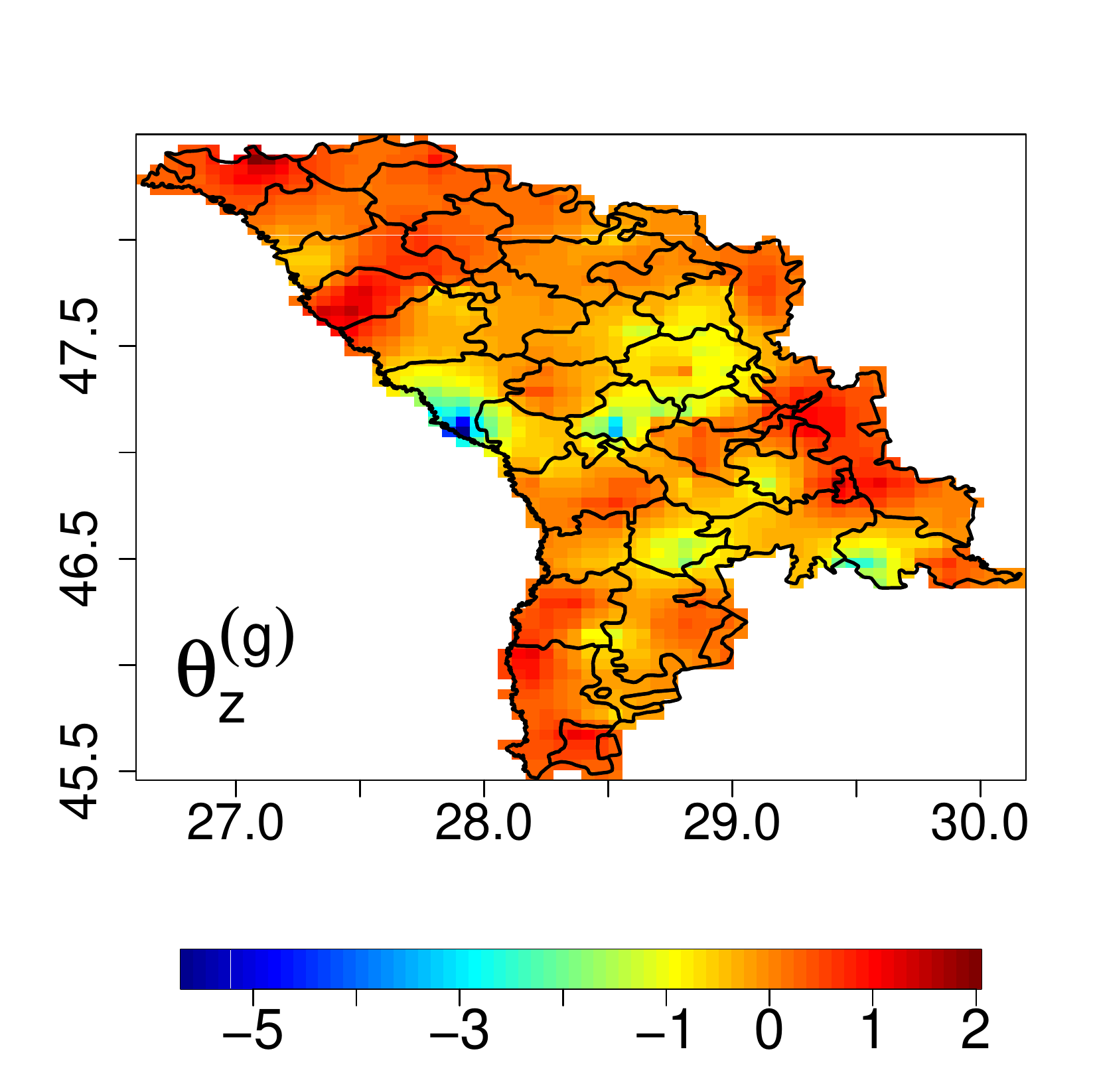}
\includegraphics[trim={0.0cm 0.0cm 0.0cm 1.0cm}, clip, scale = 0.30]{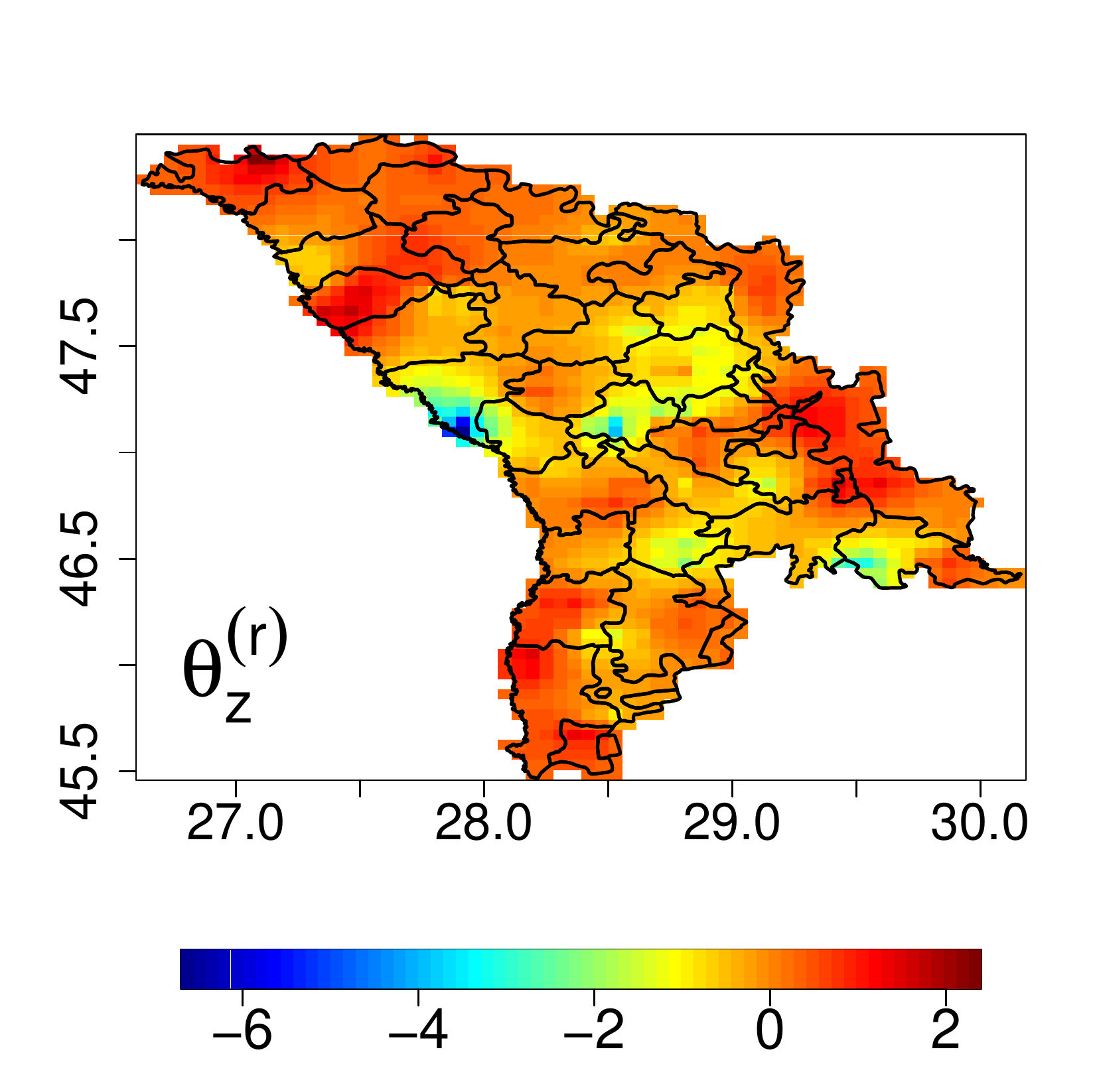}\\
\includegraphics[trim={0.0cm 0.0cm 0.0cm 1.0cm}, clip, scale = 0.30]{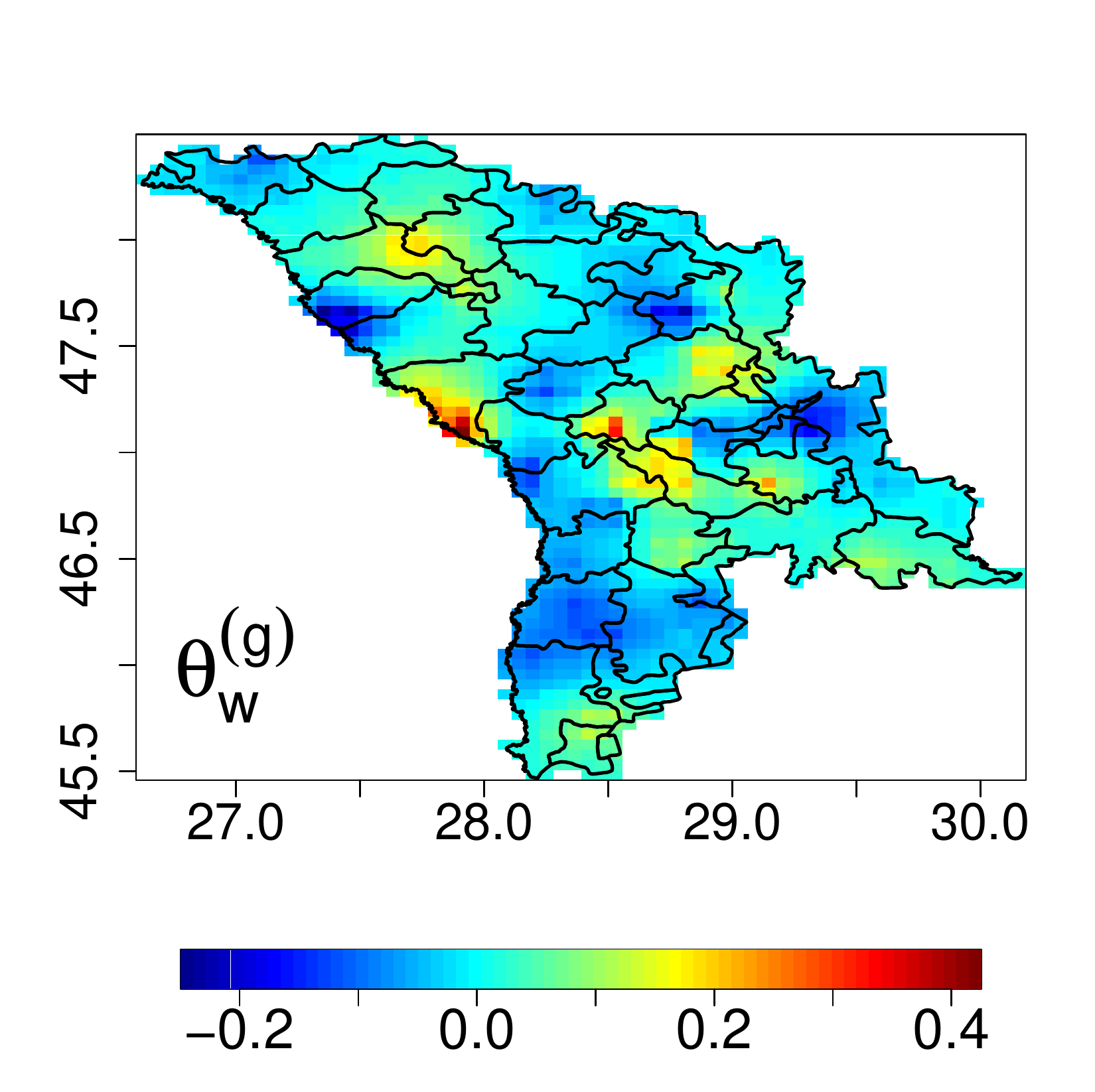}
\includegraphics[trim={0.0cm 0.0cm 0.0cm 1.0cm}, clip, scale = 0.30]{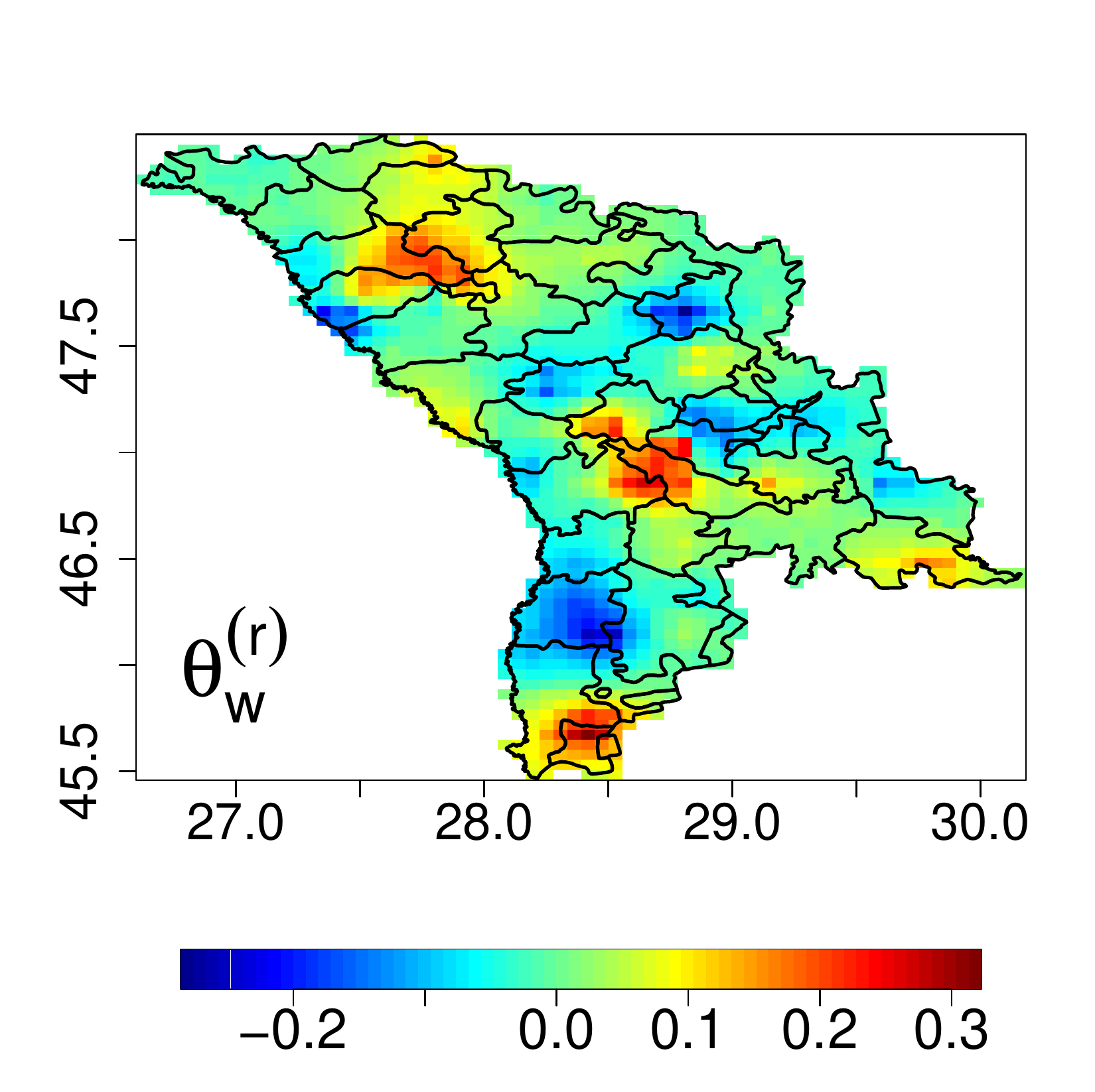}
\caption{Predicted random effect parameter results (posterior means) from the Republic of Moldova data analyses.  Patristic distance results (top row); transmission probability results (rows two and three).  $z$: binary component; $w$: continuous component; $g$: giver role; $r$: receiver role.}
\end{figure}
\clearpage 

\begin{figure}[ht]
\centering
\includegraphics[trim={0.0cm 0.0cm 0.0cm 1.0cm}, clip, scale = 0.30]{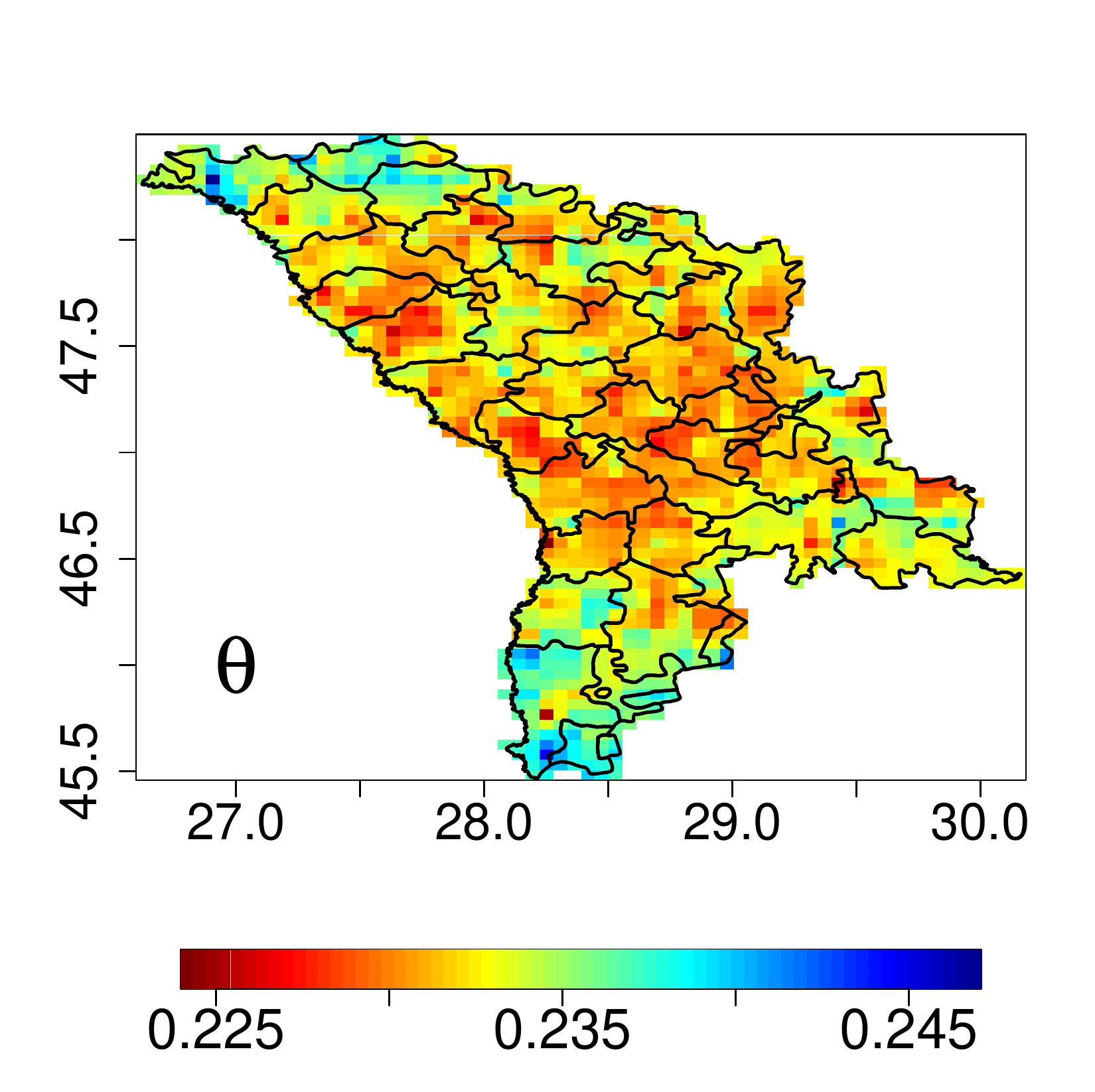}\\
\includegraphics[trim={0.0cm 0.0cm 0.0cm 1.0cm}, clip, scale = 0.30]{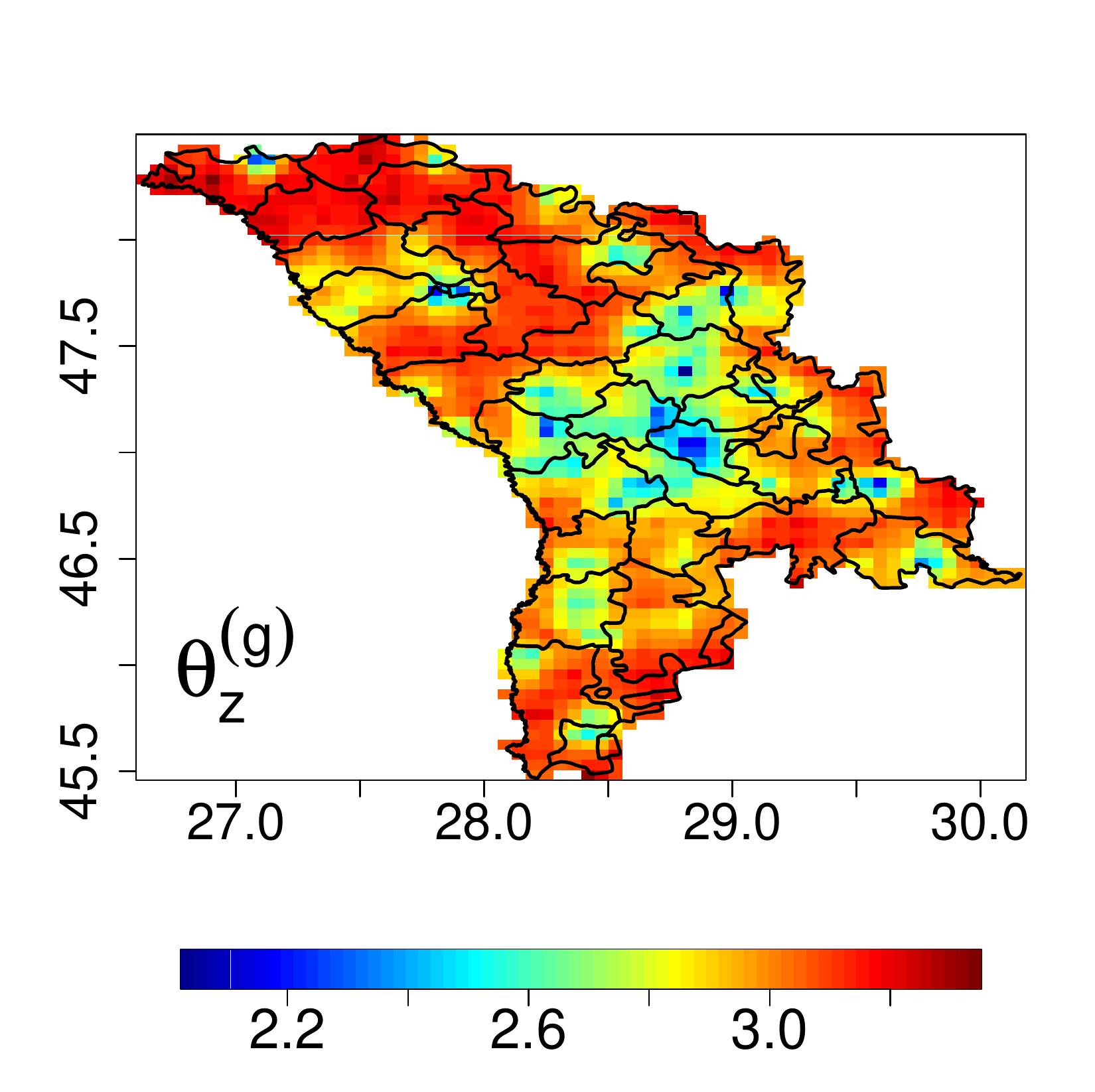}
\includegraphics[trim={0.0cm 0.0cm 0.0cm 1.0cm}, clip, scale = 0.30]{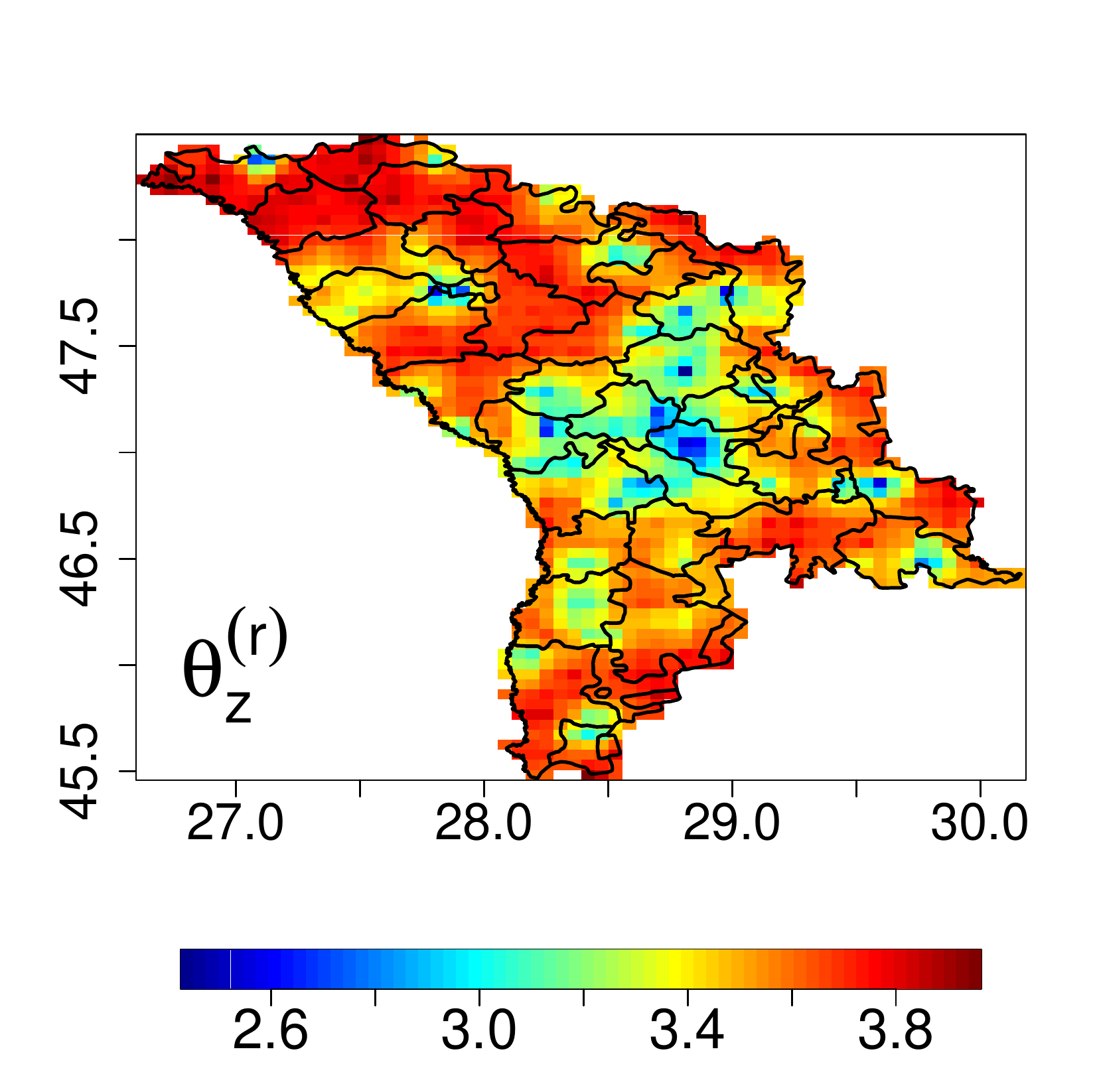}\\
\includegraphics[trim={0.0cm 0.0cm 0.0cm 1.0cm}, clip, scale = 0.30]{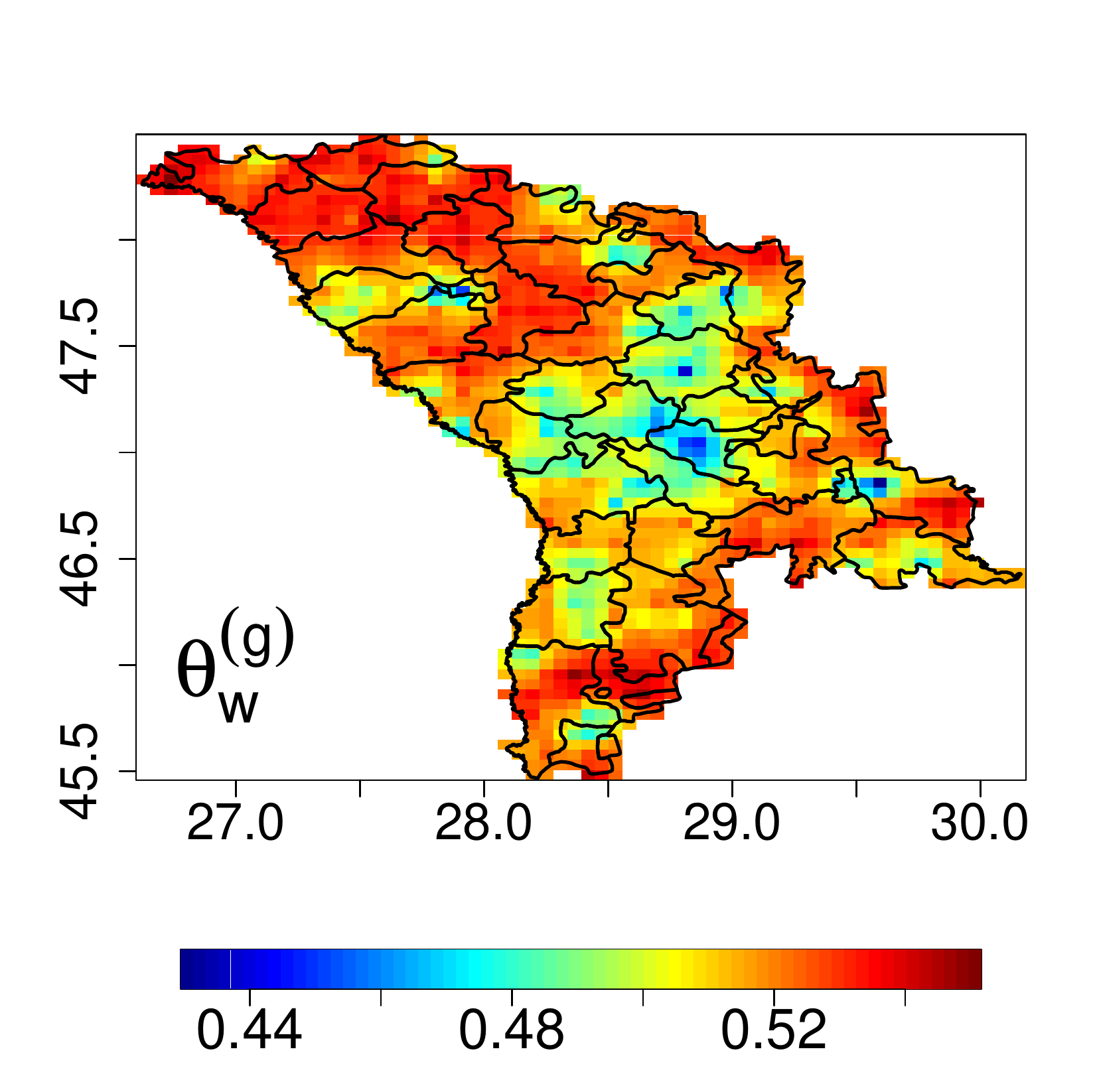}
\includegraphics[trim={0.0cm 0.0cm 0.0cm 1.0cm}, clip, scale = 0.30]{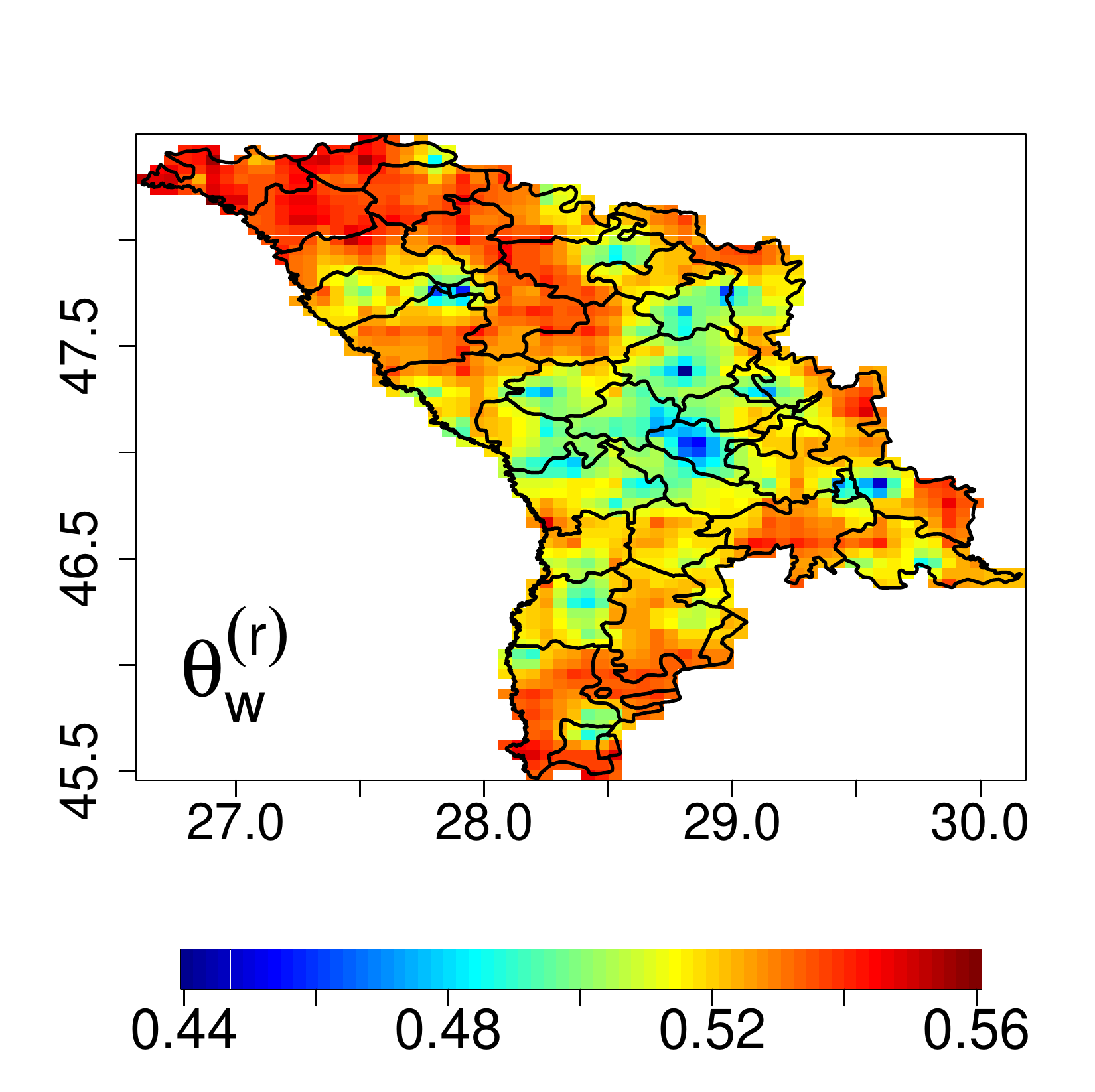}
\caption{Predicted random effect parameter results (posterior standard deviations) from the Republic of Moldova data analyses.  Patristic distance results (top row); transmission probability results (rows two and three).  $z$: binary component; $w$: continuous component; $g$: giver role; $r$: receiver role.}
\end{figure}
\clearpage

\begin{landscape}
\begin{table}[ht!]
\centering
\caption{Simulation study true parameter values obtained from the Republic of Moldova data analyses (posterior means).  PD: patristic distances; TP: transmission probabilities; $z$: binary component; $w$: continuous component; $g$: giver role; $r$: receiver role.}
\begin{tabular}{lrrrrr}
\hline
Parameter                                           & PD        & TP $(z, g)$                 & TP $(z, r)$ & TP $(w, g)$                 & TP $(w, r)$ \\
\hline
Intercept                                           & -8.1265   & \multicolumn{2}{c}{-5.1612}               & \multicolumn{2}{c}{-7.6490}               \\
Distance Between Villages                           &  0.0026   & \multicolumn{2}{c}{-0.0153}               & \multicolumn{2}{c}{-0.0785}               \\
Same Village                                        & -0.5921   & \multicolumn{2}{c}{ 1.6007}               & \multicolumn{2}{c}{ 2.3335}               \\
Date of Diagnosis Distance                          &  0.0039   & \multicolumn{2}{c}{-0.2310}               & \multicolumn{2}{c}{-0.2468}               \\
Age Difference                                      &  0.0003   & \multicolumn{2}{c}{ 0.0237}               & \multicolumn{2}{c}{-0.0697}               \\
Age                                                 & -0.0088   &  0.2840                     &  0.2875     & -0.0707                     & -0.0058     \\
Sex (Male vs.\ Female):                             &           & -0.2443                     & -0.4785     & 0.1283                      & 0.1726      \\
\ \ \ Mixed Pair vs.\ Both Female                   & -0.0171   &                             &             & & \\
\ \ \ Both Male vs.\ Both Female                    & -0.0215   &                             &             & & \\
Residence Location (Urban vs.\ Rural):              &           & -1.1754                     & -1.5636     & 0.1929                      & 0.2231      \\
\ \ \ Mixed Pair vs.\ Both Rural                    & -0.0224   &                             &             & &  \\
\ \ \ Both Urban vs.\ Both Rural                    & -0.0348   &                             &             & & \\
Working Status (Employed vs.\ Unemployed):          &           &  1.3725                     &  1.5920     & 0.0264                      & 0.1653      \\
\ \ \ Mixed Pair vs.\ Both Unemployed               & -0.0216   &                             &             & & \\
\ \ \ Both Employed vs.\ Both Unemployed            & -0.0170   &                             &             &                             &             \\
Education ($<$ Secondary vs.\ $\geq$ Secondary):    &           &  1.0456                     &  1.2273     & 0.0684                      & 0.1255      \\
\ \ \ Mixed Pair vs.\ Both $\geq$ Secondary         &  0.0503   &                             &             & & \\
\ \ \ Both $<$ Secondary vs.\ Both $\geq$ Secondary &  0.1058   &                             &             & & \\
\hline 
$\sigma^2_{\epsilon}$                               &  0.0548   &                             &             & \multicolumn{2}{c}{2.3799}                \\
$\sigma^2_{\nu_z}$                                  &           & \multicolumn{2}{c}{7.4912}                & &                                         \\
$\sigma^2_{\nu_w}$                                  &           & &                                         & \multicolumn{2}{c}{1.5168}               \\
$\tau^2 + \sigma^2_{\zeta}$                         &  0.0784   &                             &             &                             &             \\
$\Omega_{11} + \sigma^2_{\zeta^{(g)}_{z}}$          &           & 10.6231 &  & &                                                         \\
$\Omega_{22} + \sigma^2_{\zeta^{(r)}_{z}}$          &           & & 14.9655  & &                                                         \\
$\Omega_{33} + \sigma^2_{\zeta^{(g)}_{w}}$          &           & & & 0.2945 &                                                           \\
$\Omega_{44} + \sigma^2_{\zeta^{(r)}_{w}}$          &           & & & & 0.2958                                                           \\
$\Omega_{12}$                                       &           & \multicolumn{4}{c}{ 12.3052}                                                           \\
$\Omega_{13}$                                       &           & \multicolumn{4}{c}{-0.7456}                                                           \\
$\Omega_{14}$                                       &           & \multicolumn{4}{c}{-0.1850}                                                           \\
$\Omega_{23}$                                       &           & \multicolumn{4}{c}{-0.8779}                                                           \\
$\Omega_{24}$                                       &           & \multicolumn{4}{c}{-0.2136}                                                           \\
$\Omega_{34}$                                       &           & \multicolumn{4}{c}{ 0.1324}                                                           \\
\hline
\end{tabular}
\end{table}
\end{landscape}
\clearpage

\begin{table}[ht!]
\centering
\begin{threeparttable}[b]
\caption{Random effect parameter results from all analyses in the Republic of Moldova.  The table presents the number of parameters with 95\%  quantile-based equal-tailed credible intervals (CIs) that exclude zero (positive and negative), summaries of the outcome variables (i.e., patristic distances and transmission probabilities (TP)) overall and for each significance group, and the proportion of total variability in the parameters that is spatially structured.}
\begin{tabular}{lrrrrrrrr}
\hline
                         & \multicolumn{2}{c}{\# of Significant $\theta_i$} & & \multicolumn{3}{c}{Outcome Summary\footnotemark[1]} & &                        \\
\cline{2-3} \cline{5-7}
Model                    & $> 0$                            & $< 0$ & & Overall & $\theta_i > 0$ & $\theta_i < 0$             & & Spatial Variability\footnotemark[2] \\
\hline
Patristic                & 29                               & 28    & & 3.12    & 3.96  & 2.47              & & 0.23 (0.05, 0.68)      \\
TP: Binary, Giver        & 45                               & 26    & & 71.72   & 70.74 & 73.41             & & 0.98 (0.96, 1.00)                       \\
TP: Binary, Receiver     & 44                               & 29    & & 71.72   & 70.70 & 73.57             & & 0.98 (0.96, 0.99)      \\
TP: Continuous, Giver    & 9                                & 7     & & 0.86    & 2.29  & 0.42              & & 0.74 (0.50, 0.93)      \\
TP: Continuous, Receiver & 8                                & 6     & & 0.86    & 0.75  & 0.65              & & 0.76 (0.49, 0.95)      \\
\hline 
\end{tabular}
  \begin{tablenotes}
    \item[1] Patristic: Average patristic distance multiplied by 10,000; TP: Binary:  Percentage of transmission probabilities equal to zero; TP: Continuous:  Average of the positive transmission probabilities multiplied by 100.  
    \item[2] Patristic:  Posterior mean and 95\% CI of $\tau^2/(\tau^2 + \sigma^2_{\zeta})$; TP: Binary, Giver:  Posterior mean and 95\% CI for $\omega_{11}/(\omega_{11} + \sigma^2_{\zeta_z^{(g)}})$ where $\omega_{11}$ is the $(1,1)$ entry of $\Omega$; similar definitions for TP: Binary, Receiver; TP: Continuous, Giver; and TP: Continuous, Receiver. 
  \end{tablenotes}
 \end{threeparttable}
\end{table}
\clearpage 
\bibliographystyle{chicago}
\bibliography{references.bib}